\documentclass[hyper]{JHEP3}
\usepackage{graphicx}

\input{epsf}
\usepackage{epsfig}
\usepackage{amssymb}
\usepackage{amsfonts}
\usepackage{amsbsy}
\usepackage[all]{xy}
\usepackage{amsmath}

\usepackage{amssymb,amscd}
\usepackage{mathrsfs}
\usepackage{amsmath,amsthm}

\def\mod{{\rm mod}}
\def\Tr{{\rm Tr}}

\def\IC{\mathbb{C}}

\def\IH{\mathbb{H}}

\def\IZ{{\mathbb{Z}}}

\def\CC {{\cal C}}

\def\CN {{\cal N}}
\def\CR {{\cal R}}
\def\CQ {{\cal Q}}

\def\CF {{\cal F}}

\def\CP {{\cal P }}

\def\CO {{\cal O}}

\def\CH {{\cal H}}

\def\half{\frac{1}{2}}

\def\one{{\hbox{ 1\kern-.8mm l}}}
\def\tr{{\rm tr\,}}

\def\half{\frac{1}{2}}

\newcommand\be{\begin{equation}}
\newcommand\ee{\end{equation}}

\newcommand{\ie}{{\it i.e.~}}

\title{Extremal $\CN=(2,2)$ 2D Conformal Field Theories and
  Constraints of Modularity }

\author{Matthias R.~Gaberdiel,$^1$ Sergei Gukov,$^{2,3,}$\footnote{On
leave from California Institute of Technology.}~ Christoph A.~Keller,$^1$
Gregory~ W.~Moore,$^4$ and Hirosi ~Ooguri$^{5,6}$
\\ ~
\\
$^1$
Institut f\"ur Theoretische Physik, ETH Zurich,
CH-8093 Z\"urich, Switzerland \\
$^2$ School of Mathematics,\\
~ Institute for Advanced Study, Princeton, NJ 08540, USA\\
$^3$ Department of Physics and Department of Mathematics, \\
~ University of California, Santa Barbara, CA 93106, USA\\
$^4$ NHETC and Department of Physics and Astronomy,\\
~ Rutgers University, Piscataway, NJ 08855--0849, USA\\
$^5$ California Institute of Technology, Pasadena, CA 91125, USA \\
$^6$ Institute for the Physics and Mathematics of the Universe,\\
~ University of Tokyo,
Kashiwa, Chiba 277-8586, Japan}

\abstract{We explore the constraints on the spectrum of primary
fields implied by   modularity of the elliptic genus of $\CN=(2,2)$
2D CFT's. We show that such constraints have nontrivial implications
for the existence of ``extremal'' $\CN=(2,2)$ conformal field
theories. Applications to $AdS_3$ supergravity and flux
compactifications are addressed.
\\
\\
\\
\\
\\
\\
{\tt CALT-68-2685, IPMU 08-0031, ITEP-TH-20/08}}

\begin{document}


\section{Introduction and summary}

In recent work \cite{Witten:2007kt} Witten has revived the study of
$2+1$ dimensional quantum gravity. In particular, he has defined a
notion of pure $AdS_3$ quantum gravity and investigated its
properties in light of the AdS/CFT correspondence.  These
considerations naturally lead to a notion of an \emph{extremal
conformal field theory.} Extremality means that the partition
function of the boundary CFT is as close as possible to the Virasoro
character of the vacuum. The reason for this is that there are two
kinds of excitations in pure gravity: the perturbative excitations
and the black holes. The perturbative excitations are identified
with Virasoro descendants of the vacuum following
\cite{Brown:1986nw} while the   Virasoro primaries correspond to the
BTZ black holes. Since black holes are parametrically heavy there is
a large gap from the vacuum to the first nontrivial Virasoro
primary.  The present paper addresses similar questions for pure
quantum gravity with extended $\CN=2$ supersymmetry. Our main tool
will be the elliptic genus of an $\CN=2$ superconformal field
theory. As we recall below, this is a weak Jacobi form, and its
modular properties impose tight constraints on the partition
function. The advantage of this approach is that, unlike the case of
\cite{Witten:2007kt}, we do not have to assume the complete
holomorphic factorization of the partition function. The holomorphy
and modularity of the elliptic genus holds for any conformal field
theory with $\CN=2$ supersymmetry. Thus, we can test the
hypothetical existence of a theory of pure $AdS_3$ supergravity
  without relying on the additional assumption of
holomorphic factorization. We will show that there is some tension
between these modular properties and the notion of extremality.

A brief summary of our results is the following:

\begin{enumerate}

\item  In  section \ref{secdefinition} we give a   definition of an
extremal $(2,2)$ superconformal field theory   which, one might
expect would constitute a holographic dual to ``pure $(2,2)$ $AdS_3$
supergravity.'' In any case, it is a natural generalization of the
notion of extremality to  $(2,2)$ supersymmetry. In this paper we
restrict attention to theories with $c=0\mod 6$ and integral
$U(1)_R$ charges for the left- and right-moving $\CN=2$ algebras.
Relaxing this assumption is an interesting open problem.

\item In section \ref{sec:SearchForEG} we give numerical evidence
that only a finite number of ``exceptional'' examples of extremal
$(2,2)$ theories can exist. Then in section \ref{sec:SearchAnalytic}
we give an analytic proof that this is indeed the case. We also
present very strong evidence that the extremal elliptic genus only
exists for nine values of $c$, namely
\be 6,12,18,24,30,42,48,66,78 .\ee

\item  In section \ref{sec:NearExtremalCFT} we
then introduce the notion of a ``nearly extremal $(2,2)$
superconformal theory,'' whose spectrum only approximates that of
pure $(2,2)$ supergravity. We show that if the degree of
approximation is relaxed then candidate elliptic genera do indeed
exist.

\item   By quantifying the degree of approximation required to produce
candidate elliptic genera we are able to constrain the spectrum as
follows. Consider states (in the NSNS sector) which are
right-chiral-primary and left $\CN=2$ primary with $(L_0,J_0)$
eigenvalue $(h,\ell)$. In
section \ref{sec:modconstspec}, equation (\ref{bound}) we show that
for $c$ large any theory with modular elliptic genus must have some
such state with
\be   h  < \frac{c}{24} + \frac{3 \ell^2}{2c} -\frac{1}{8} +
\CO(c^{-1/2}) \ .
\ee
This result is conjectural. It is supported by numerical evidence
described in section \ref{sec:NearExtremalCFT}. Finding a rigorous
justification of (\ref{bound}) (or a counterexample) is an
interesting open problem raised by the present paper.

\item On the   other hand, in section \ref{sec:ConstructionNearlyExtremal}
we show that it is possible to construct an elliptic genus which is
compatible with the spectrum of an extremal $(2,2)$ superconformal
theory for conformal weights $h \leq \frac{c}{24}$.

\item In section  \ref{sec:ExtremeN=4} we comment on a partial
generalization of our results
to $\CN=4$ theories.

\end{enumerate}

In the remainder of the paper we discuss some implications of the
above results.  First, in section \ref{sec:Discussion} we discuss
the implications for the existence of pure $(2,2)$ $AdS_3$
supergravity. While our results cast some doubt on the existence of
such theories, they are not conclusive. It is conceivable that
quantum corrections to the cosmic censorship bound for the existence
of black holes imply that one should identify a near-extremal rather
than an extremal $(2,2)$ CFT as a holographic dual of pure
supergravity. We leave this question for future work. Of course,
even when a candidate Jacobi form exists that does not mean a
corresponding $(2,2)$ supergravity necessarily exists. In the
analogous  $\CN=0$ case the relevant partition functions can readily
be constructed, but it is not known whether the corresponding
extremal CFT's exist for general Chern-Simons levels $k$. Indeed,
there is an argument based on the modular differential equation of
these partition functions \cite{Gaberdiel:2007ve,Gaberdiel:2008pr}
that suggests that the theories are in fact inconsistent for
sufficiently large $k$.

A second motivation for the present work is that constraints on
conformal field theory spectra implied by modular invariance might
have interesting applications to flux compactifications of string
theory and M-theory. This is briefly explained in section
\ref{sec:AppFluxCompt}. Again, the development of this idea is left
to future work.


\section{Polar states and the elliptic genus}
\label{sec:PolarStatesEG}

We will focus on theories with $\CN=(2,2)$ two-dimensional
superconformal symmetry.   It will be convenient to parametrize the
(left = right) central charge as $c=6m$. A simple example of such a
theory that the reader might wish to keep in mind is an $\CN=(2,2)$
sigma-model based on a Calabi-Yau target space of complex dimension
$2m$. In the present paper we  only consider integer values of $m$,
and thus the relevant Calabi-Yau manifolds have even complex
dimension.\footnote{A generalization to half-integer values of $m$
should be possible, but we will not attempt it in the present paper.
For $m=1$ the resulting theory actually has $(4,4)$ supersymmetry,
but we will not use this fact.  } In particular, the smallest
non-trivial value of $m$ corresponds to a Calabi-Yau 2-fold, that is
a torus $T^4$ or a $K3$ surface.

We assume that the Hilbert space of our theory is a direct sum of
unitary highest weight representations of the $\CN=2$ algebra. This allows
us to define the RR-sector partition function
\begin{equation}\label{eq:NSNSpf}
 Z_{RR}(\tau,z;\bar\tau,\bar z):= {\Tr}_{\CH_{RR}}
q^{L_0-c/24} e^{2\pi i z J_0}\bar q^{\tilde L_0-c/24} e^{2\pi i \bar
z \tilde J_0}e^{i \pi (J_0 - \tilde J_0)}
\end{equation}
which has good modular properties under the $SL(2,\IZ)$ action
$(\tau,z) \to (\frac{a \tau + b}{c \tau + d}, \frac{z}{c\tau +d} )$. Here,
as usual, $q=e^{2\pi i \tau}$ and $y=e^{2\pi i z}$, and similarly for
$\bar{q}$ and $\bar{y}$.

In these conventions, the elliptic genus of an $\CN = (2,2)$
superconformal field theory $\CC$  is defined to be
\begin{equation}\label{eq:DefEllipGen}
\chi(\tau,z;\CC):= Z_{RR}(\tau,z  ;\bar \tau, 0) \ .
\end{equation}
It is holomorphic in $(\tau,z)$ by the standard properties of the
Witten index. For references on the elliptic genus see
\cite{AlvarezWG,AlvarezDE,EOTY,Kawai:1993jk,LercheQK,LercheCA,
O,PilchEN,PilchGS,SchellekensYJ,SchellekensYI,WindeyAR,witteneg}.
\smallskip

$\CN=2$ algebras have the crucial spectral flow isomorphism
\cite{Schwimmer:1986mf}, which allows us to relate the NS and
R-sector partition functions. Recall that spectral flow
$SF_{\theta}$ for $\theta \in \half \IZ$ is an isomorphism of
$\CN=2$ superconformal algebras which maps eigenvalues
\begin{eqnarray}
L_0 & \rightarrow & L_0 + \theta J_0 + \theta^2 m \\
J_0 & \rightarrow & J_0 + 2\theta m \ .
\end{eqnarray}
The spectral flow operators act on $Z=Z_{RR}$ as:
\be (SF_\theta \widetilde{SF}_{\tilde \theta} Z) =  e\left(m
\theta^2 \tau + 2m\theta(z+\half)\right) \cdot  e\left(m
\tilde\theta^2 \bar \tau + 2m\tilde \theta(\bar z-\half)\right)
Z(\tau,z+\tau\theta;\bar \tau, \bar z + \tilde \theta \bar \tau),
\ee
where $e(x):=e^{2\pi i x}$.  For simplicity we   restrict our
attention to theories with integral spectrum of left- and
right-moving $U(1)$ charges $J_0,\tilde J_0$. Again, it should be
possible, and would be interesting, to relax this assumption.
Spectral-flow invariant theories with integral $U(1)$ charges
satisfy
\begin{eqnarray}
Z_{RR} & = &  (SF_\theta \widetilde{SF}_{\tilde \theta}) Z_{RR}
\qquad
\theta, \tilde\theta \in \IZ \\
Z_{NSNS} & = &  (SF_\theta \widetilde{SF}_{\tilde \theta}) Z_{RR}
\qquad
\theta, \tilde\theta \in \IZ+\half \ .
\end{eqnarray}

As is well-known \cite{Kawai:1993jk}, the  modularity properties of
$Z_{RR}$ together with spectral flow invariance and unitarity imply
that the elliptic genus is a \emph{weak Jacobi form} of index $m$
and weight zero \cite{EichlerZagier}. A weak Jacobi form
$\phi(\tau,z)$ of weight $w$ and index $m\in{\mathbb Z}$, with
$(\tau,z)\in \IH\times \IC$, satisfies the transformation laws
\begin{equation}\label{eq:jactmn1}
 \phi({a \tau + b \over  c \tau + d} , {z \over  c \tau + d}) =
(c \tau+d)^w e^{ 2 \pi i m { c z^2 \over  c \tau + d} } \phi(\tau,z)
\qquad \begin{pmatrix} a & b \\ c & d \\ \end{pmatrix} \in SL(2,\IZ) \ ,
\end{equation}
\begin{equation}\label{eq:jactmn2}
 \phi(\tau,z+ \ell \tau + \ell') = e^{-2 \pi i m
(\ell^2 \tau+ 2 \ell z)} \phi(\tau,z) \qquad\qquad\qquad
\ell,\ell'\in \IZ \ , \end{equation}
and has a Fourier expansion
\begin{equation}
 \phi(\tau,z) = \sum_{n \geq 0, \ell\in \IZ} c(n,\ell) q^n
y^\ell \end{equation}
with $c(n,\ell) = (-1)^w c(n,-\ell)$. It follows
from the spectral flow identity that $c(n,\ell)=0$ for $4mn-\ell^2<-m^2$.
Following \cite{EichlerZagier}, we denote by $\tilde J_{w,m}$
the vector space of weak Jacobi forms of weight $w$ and  index $m$.
A Jacobi form is then a weak Jacobi form whose polar part
vanishes (see below).

Suppose we are given an integer $m\in \IZ_+$. If $(\ell,n)\in \IZ^2$
is a lattice point we refer to its \emph{polarity} as
$p=4mn-\ell^2$.  If $\phi \in \tilde J_{0,m}$ let   us define the
\textit{polar part} of $\phi$, denoted $\phi^-$, to be the sum of the
 terms in the Fourier expansion corresponding to lattice points of
 negative polarity.   By spectral flow one can always relate
the degeneracies to those in the fundamental domain with $\vert
\ell\vert \leq m$. If we impose the modular transformation
(\ref{eq:jactmn1}) with $-\mathbf{1}\in SL(2,\mathbb{Z})$, which implements
charge conjugation, then $c(n,\ell) =
c(n,-\ell)$ and therefore the polar coefficients which cannot be
related to each other by spectral flow and charge conjugation are
$c(n,\ell)$ where $(\ell,n)$ is valued in the \emph{polar region}
$\CP$ (of index $m$), defined to be
\begin{equation}\label{eq:polarregion}
\CP^{(m)} := \{ (\ell,n): 1\leq \ell \leq m, \quad  0 \leq n , \quad
p=4mn-\ell^2<0 \} \ .
\end{equation}
For an example, see figure \ref{polarfig}.

\EPSFIGURE{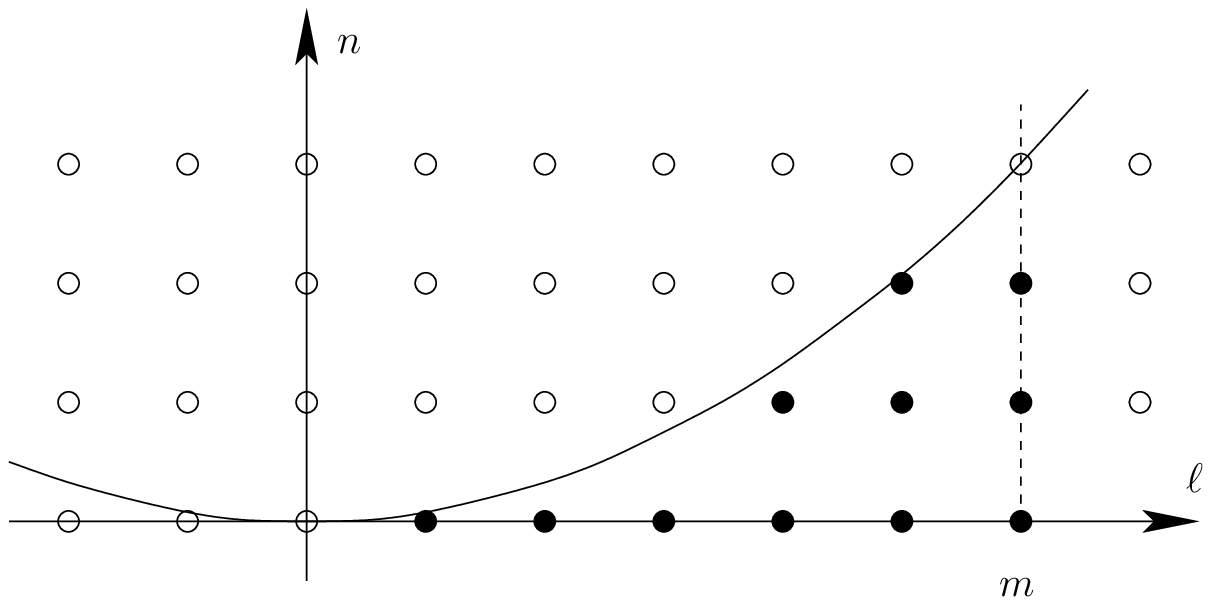,height=6cm,angle=0,trim=0 0 0 0}%
{A cartoon showing polar states (represented by ``$\bullet$'') in
the region $\CP^{(m)}$. Spectral flow by $\theta={1 \over 2}$
relates these states to particle states in the NS sector of an
$\CN=2$ superconformal field theory which are holographically dual
to particle states in $AdS_3$.
\label{polarfig} }


  Given any Fourier expansion
\begin{equation}
\psi(\tau,z) = \sum_{\ell,n\in \IZ}  \hat \psi(n,\ell) q^n y^\ell
\end{equation}
we define its  \emph{ polar polynomial} (of index $m$) to be the sum
restricted to the polar region $\CP^{(m)}$:
\begin{equation}
{\rm Pol}(\psi) := \sum_{(\ell,n)\in \CP^{(m)}} \hat \psi(n,\ell)
q^n y^\ell \ .
\end{equation}
Let us moreover denote by $V_m$ the space of polar polynomials, \ie
the vector space generated by the monomials $q^n y^\ell$ with
$(\ell,n)\in \CP^{(m)}$.

The key mathematical fact we need follows from the theory of
``periods of modular forms.'' The upshot is that one can reconstruct
a weak Jacobi form of weight zero from its polar polynomial.
Moreover there is a sequence
\begin{equation}\label{eq:exactseq}
0 \rightarrow \tilde J_{0,m} ~{\buildrel {\rm Pol} \over
\rightarrow}~  V_m ~{\buildrel {\rm Per} \over \rightarrow}~ S_{5/2}
\end{equation}
exact at $V_m$, where ${\rm Per}$ is a ``period map'' to a certain
space of vector-valued cusp forms of weight $5/2$.   A nonzero image
in the space of cusp forms means that the polar polynomial cannot be
realized by a true weak Jacobi form. For an explanation of these
facts in the physics literature, together with references to the
mathematical literature, see \cite{Dijkgraaf:2000fq,Moore:2004fg,
Manschot:2007ha,Manschot:2008zb}. The reader interested in these
matters should also consult \cite{Borcherds}.

In the next two sections we will show that there can indeed be
nontrivial obstructions simply by computing the dimensions of
$\tilde J_{0,m}$ and $V_m$.

Returning to the conformal field theory $\CC$, an  eigenstate of
$L_0,J_0$ is called a  \textit{polar state} if it has negative
polarity:
\begin{equation}
p=4mL_0 - J_0^2-m^2= 4m (L_0 - \frac{c}{24}) - J_0^2 <0 \ .
\end{equation}
One checks that $4mL_0-J_0^2$ is spectral flow invariant, so we can
speak of polar states in both the R and NS sector. Using the
mathematical results explained above we see that the   significance
of polar states is that the polar degeneracies of the elliptic genus
determine all the other Fourier coefficients of the elliptic genus.

\subsection{Counting weight zero weak Jacobi forms}

Let $\tilde J_{ev,*} = \oplus_{w\in 2\IZ, m\in \IZ} \tilde J_{w,m}$
denote the bigraded ring of weak Jacobi forms of even weight.
According to \cite{EichlerZagier}, Theorem 9.3, $\tilde J_{ev,*}$ is
a polynomial algebra on four generators of degree \be
\label{degjacobi} (w,m) ~=~(4,0),~~ (6,0),~~ (-2,1),~~ (0,1)\ . \ee
The first two generators correspond to the Eisenstein series
\be \label{efour}
E_4 = 1 + 240 \sum_{n=1}^{\infty} \sigma_3 (n) q^{n}
= 1 + 240 q + 2160 q^2 + 6720 q^3 + \ldots
\ee
 and
\be \label{esix}
E_6 = 1 - 504 \sum_{n=1}^{\infty} \sigma_5 (n) q^{n} =
1 - 504 q - 16632 q^2 - 122976 q^3 - \ldots\ ,
\ee
where the
Fourier coefficients $\sigma_k(n) := \sum_{d \vert n} d^k$ are
defined to be the $k$-th powers of the positive divisors of $n$. A
generalization of Eisenstein series to Jacobi forms is described in
\cite{EichlerZagier}:
\begin{equation}
E_{k,m}(\tau,z) = \half \sum_{c,d\in \IZ, (c,d)=1} \sum_{\ell\in
\IZ} (c\tau+d)^{-k} e^{2\pi i m \left( \ell^2 \frac{a \tau + b}{c
\tau+d} + 2 \ell \frac{z}{c\tau +d}
- \frac{c z^2}{c\tau +d}\right)} \ .
\end{equation}

In terms of these generalized Eisenstein series one can write the
remaining two generators in (\ref{degjacobi}) as
 \be
\label{phiphitilde} \tilde \phi_{-2,1} = {\phi_{10,1} \over \Delta}
\in \tilde J_{-2,1} \quad\quad\quad \tilde \phi_{0,1} = {\phi_{12,1}
\over \Delta} \in \tilde J_{0,1}  \ ,
\ee
where the first subscript on
$\tilde \phi$ denotes the weight and the second denotes the index.
Here, $\Delta = q \prod_{n=1}^{\infty} (1-q^n)^{24}$ and
\be
\label{cuspgen}
\begin{split}
\phi_{10,1} & = {1 \over 144} (E_6 E_{4,1} - E_4 E_{6,1}) \\
& = (y-2+y^{-1}) q + (-2y^2 - 16y + 36 - 16y^{-1} -2 y^{-2}) q^2
+ \ldots\ , \\
\phi_{12,1} & = {1 \over 144} (E_4^2 E_{4,1} - E_6 E_{6,1}) \\
& = (y+10+y^{-1}) q + (10y^2 - 88y -132 - 88y^{-1} +10 y^{-2}) q^2
+ \ldots\ . \\
\end{split}
\ee
Thus the two weak Jacobi forms $\tilde \phi_{-2,1}$ and
$\tilde\phi_{0,1}$ have the series expansion
\be\label{phiphitildeexp}
\begin{split}
\tilde\phi_{-2,1} & = \left(y-2+y^{-1}\right)
+\left(-2 y^2+8y-12+8y^{-1}-2y^{-2}\right) q + \ldots\ , \\
\tilde\phi_{0,1} & = (y +10 + y^{-1}) +
\left(10 y^2-64y+108-64 y^{-1}+10y^{-2}\right) q + \ldots  \ .
\end{split}
\ee
Much useful information about Jacobi forms  can be found in
\cite{EichlerZagier}.

To summarize, a natural vector space basis of $\tilde J_{0,m}$ is given by
\begin{equation}\label{natbasis}
(\tilde \phi_{-2,1})^a (\tilde \phi_{0,1})^b E_4^c E_6^d\ ,
\end{equation}
where $a,b,c,d $ are nonnegative integers such that   $a+b=m$,  and
$a= 2c + 3d$. It is straightforward to compute the number of
solutions to these constraints and thereby show that
\begin{equation}
j(m):=\dim\tilde J_{0,m} = \frac{m^2}{12} + \frac{m}{2} + \biggl(
\delta_{s,0} + \frac{s}{2} - \frac{s^2}{12} \biggr)\ ,
\end{equation}
where $m = 6 \rho + s$ with $\rho\geq 0$ and $0 \leq s \leq 5$.
Specifically,
\begin{equation} \label{dim_jm}
j(m)  = \begin{cases} m^2/12 + m/2 + 1 & m=0\ \mod\ 6\\
m^2/12 + m/2  + 5/12 & m=1,5\ \mod\ 6\\
m^2/12 + m/2  + 2/3  & m=2,4\ \mod\ 6\\
m^2/12 + m/2  + 3/4 & m=3\ \mod\ 6\ .\\
\end{cases}
\end{equation}

\subsection{Counting polar monomials}
\label{subsec:CountingPolar}

Let us now compute the dimension of the space $V_m$, and compare it
to $j(m)$. In other words, we wish to count the number of integer
points in the $(\ell,n)$ plane bounded (on one side)
by the parabola $4mn - \ell^2 = 0$, as shown in figure \ref{polarfig}.
We have
\begin{equation}
P(m) := \dim V_m = \sum_{\ell =1}^m \lceil \frac{\ell^2}{4m} \rceil \ .
\end{equation}
Note that we want the \textit{ceiling} function and not the floor
function, as we include $n=0$ up to the largest $n$ with $n<
\ell^2/(4m)$ for each $\ell=1,\dots, m$.

To compute this we follow \cite{EichlerZagier} and write our sum as
a sum of three terms.
\be \begin{split} \sum_{\ell=1}^m \lceil \frac{\ell^2}{4m}   \rceil
= \sum_{\ell=1}^m \frac{\ell^2}{4m} - \sum_{\ell=1}^m
((\frac{\ell^2}{4m} )) + \half \sum_{\ell=1}^m \left( \lceil
\frac{\ell^2}{4m} \rceil - \lfloor \frac{\ell^2}{4m}
\rfloor)\right)\end{split} \ ,
\ee
where
\begin{equation}
((x)) := x - \half ( \lceil x \rceil + \lfloor x \rfloor) =
\begin{cases} 0 & x\in \IZ \\  \alpha-\half & x = n + \alpha,
0<\alpha<1 \ .\\ \end{cases}
\end{equation}
Note that $((x))$ is the sawtooth function. It is periodic of period
$1$.

Now we evaluate the three terms. The main term comes from the
elementary formula
\be \sum_{\ell=1}^m \frac{\ell^2}{4m} = \frac{m^2}{12} + \frac{m}{8}
+ \frac{1}{24} \ .
\ee

Next, note that the number of integers $\ell$ with $1\leq \ell \leq
m$ with $\ell^2=0\ \mod\ 4m$ is $\lfloor \frac{b}{2}\rfloor$ where $b$
is the largest integer with $b^2\vert m$. This follows from the
prime factorization of $m$.  Thus, we obtain:
\begin{equation}
  \sum_{\ell=1}^m \lceil \frac{\ell^2}{4m} \rceil
- \sum_{\ell=1}^m \lfloor \frac{\ell^2}{4m}
  \rfloor=  m - \lfloor \frac{b}{2} \rfloor \ .
\end{equation}

Finally we come to the most subtle term $\sum_{\ell=1}^m
((\frac{\ell^2}{4m} ))$. The numbers $((\frac{\ell^2}{4m} ))$ are,
very roughly speaking, randomly distributed between $-1/2$ and
$+1/2$, Therefore, the average will go to zero. In fact, they
roughly make a random walk so we expect a quantity on the order of
$m^{1/2}$. To be more precise    the discussion of
\cite{EichlerZagier}, pp. 122-124 shows that
$$
\sum_{\ell=1}^m ((\frac{\ell^2}{4m} ))= - \frac{1}{4}\sum_{d\vert
4m} h'(-d) + \frac{1}{2}((\frac{m}{4})) \ ,
$$
where $h'(-d)$ is the class number of a quadratic imaginary field of
discriminant $-d$ (with the exception of $d=3,4$).

Putting the three terms together we   obtain:
\begin{equation}
P(m) = \frac{m^2}{12} + \frac{5m}{8} + A(m)
\end{equation}
where $A(m)$ is the arithmetic function
\begin{equation}
A(m) = \frac{1}{4} \sum_{d\vert 4m} h'(-d) - \half \lfloor
\frac{b}{2} \rfloor - \half ((\frac{m}{4} )) + \frac{1}{24} \ .
\end{equation}

Very roughly speaking, $A(m)$ grows like $\CO(m^{1/2})$ so for large
$m$ we have
\begin{equation}\label{eq:Pmjm}
P(m)- j(m) = \frac{m}{8} + \CO(m^{1/2}) \ .
\end{equation}
The reader should be warned that we are not using the $\CO$ symbol
in its precise mathematical sense here, but rather as a heuristic
order-of-magnitude for the ``average'' value of the sub-leading term
to the linear behavior in \eqref{eq:Pmjm}. \footnote{This is a
subtle issue which, while fascinating, we believe is a distraction
from our main exposition. A theorem of Siegel states that
$\lim_{d\to \infty} \frac{\log h'(-d)}{\log d} = \half $ as $d$ runs
through discriminants of quadratic imaginary fields, but $h'(-d)$
itself does not have a simple asymptotic expansion. This follows
from its relation to the Dirichlet Series $L_d(s)$ at $s=1$. For a
discussion of these and related matters, together with their
possible applications to black holes and with references to the math
literature  see \cite{Miller:1999ag}. For a rigorous discussion of
the probability distribution of $h'(-d)$ see \cite{Barban,Elliot}. }

\begin{equation}\label{tab:RRfieldsSpheres}
\renewcommand{\arraystretch}{1.3}
\begin{tabular}{|c||c|c|c| }
\hline  $m$& $\dim \tilde J_{0,m} $& $\dim V_m$
\\
\hline $m=0$  & $ 1$ & $0$
\\
\hline $m=1$  & $1$ & $1$
\\
\hline $m=2$  & $2$ & $2$
\\
\hline $m=3$  & $3$ & $3$
\\
\hline $m=4$  & $4$ & $4 $
\\
\hline $m=5$  & $5 $ & $6$
\\
\hline $m=6$  & $7$ & $8  $
\\
\hline $m=7$  & $8$ & $9  $
\\
\hline $m=8$  & $10 $ & $11 $
\\
\hline $m=9$  & $12 $ & $13 $
\\
\hline $m=10$  & $14 $ & $16 $
\\
\hline $m=11$  & $16 $ & $18 $
\\
\hline $m=12$  & $19 $ & $21 $
\\
\hline
\end{tabular}
\end{equation}

The first few values of $P(m)$ and $j(m)$ are shown in the above
table. Note that $P(m)>j(m)$ for $m\geq 5$, and it is
straightforward to check with a computer that $(P(m)-j(m)-
\frac{m}{8})m^{-1/2}$ is positive and  roughly order one for  values
of $m$ out to order several thousand. See figure 2 below.

\EPSFIGURE{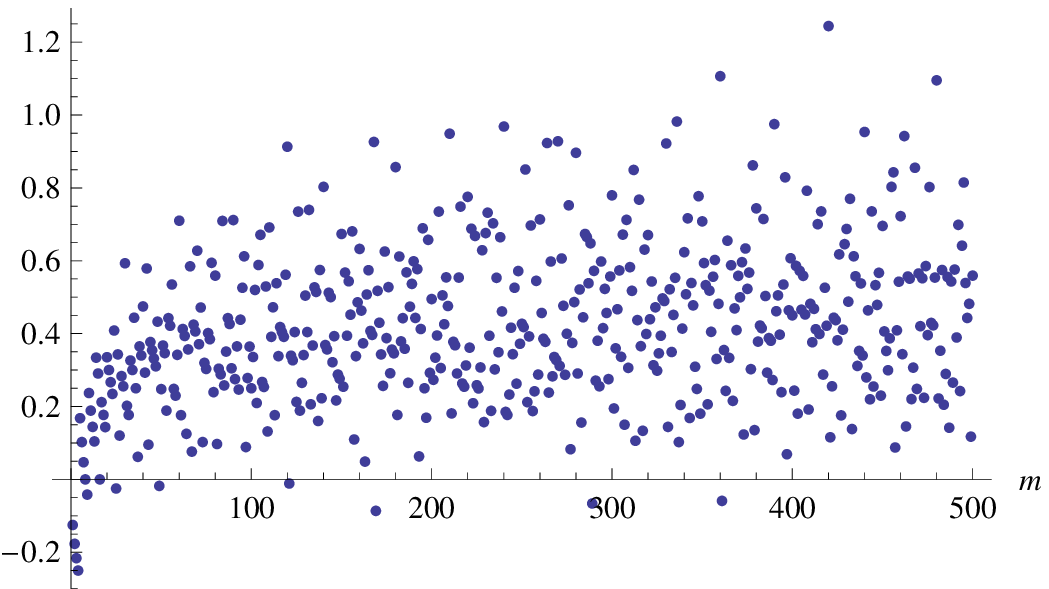,height=7cm,angle=0,trim=0 0 0 0}%
{A plot of the first few hundred values of $(P(m) -
j(m)-m/8)m^{-1/2}$ shows the quantity remains on the order of 1. The
points do not tend to a limiting value - as would be the case for an
asymptotic expansion, but are distributed about a mean value and
exhibit  a considerable amount of scatter. Detailed results on the
distribution are available in the math literature, but we will not
need these. \label{pminusjfig} }

The important conclusion that we draw is that for large $m$ there are
on the order of $\frac{m}{8}$ linear constraints on the polar coefficients
of the elliptic genus expressing modularity.

\bigskip
\textbf{Remarks}

\begin{enumerate}

\item The action of charge conjugation together with spectral flow
defines an action of $D_\infty$ on the $(\ell,n)$ plane which
preserves the space $\CQ$ of polar values $-m^2\leq 4mn-\ell^2<0$. A
fundamental domain is given by the polar region $\CP^{(m)}$, but the
quotient $\CQ/D_\infty$ has fixed points: for $\ell=-m$ the spectral
flow to $\ell=+m$ can be undone by charge conjugation. Therefore, if
we compute the \emph{orbifold} Euler character of $\CQ/D_\infty$ the
line of states $(\ell, h)$ with $\ell=m$ should be counted with
weight $\half$. There are precisely $m/4$ states on this line and
hence  $\chi_{\rm orb}(\CQ/D_\infty) = P(m)-m/8$, which is a much
closer approximation to $j(m)$.

\item Recently, J. Manschot \cite{Manschot:2008zb}  has reproduced
the formula for $P(m)-j(m)$ by directly computing the dimension of
the image of the period map ${\rm Per}$ in (\ref{eq:exactseq}).

\end{enumerate}


\section{Extremal $\CN=(2,2)$ conformal field theories}
\label{sec:ExtremalN22}

\subsection{Definition}
\label{secdefinition}

In \cite{Witten:2007kt} Witten suggested that the holographic dual
of  pure 2+1 dimensional quantum gravity should be an ``extremal
conformal field theory.'' The latter is defined to be a conformal
field theory whose modular invariant partition function is ``as
close as possible'' to the Virasoro character of the vacuum. When
$c=24k$ the vacuum character is
\begin{equation}
\chi_{Vac}^{(k)}(\tau) = q^{-k} \prod_{n=2}^\infty \frac{1}{1-q^n} \ .
\end{equation}
The partition function $Z_k(\tau)$ has weight zero. Unlike the
elliptic genus case, there is no obstruction to completing an
arbitrary polynomial in $q^{-1}$ to a modular function by adding
nonpolar terms. Therefore, Witten defines $Z_k(\tau)$ to be the
unique modular function with no singularities for $\tau \in \IH$
such that the expansion around the cusp at infinity satisfies
\begin{equation}
Z_k(\tau) :=\biggl[ q^{-k} \prod_{n=2}^\infty
\frac{1}{1-q^n}\biggr]_{q^{\leq 0}} + \CO(q) \ .
\end{equation}
Following \cite{Dijkgraaf:2000fq}, Witten interprets
 the first Virasoro primary above the vacuum
representation to be a state corresponding  to the lightest possible
BTZ black hole in $AdS_3$.

Following Witten \cite{Witten:2007kt} let us consider ``pure
$\CN=(2,2)$ supergravity'' with negative cosmological constant. This
is the hypothetical quantum theory whose classical action is a
supersymmetric completion of the Einstein-Hilbert action, \be
\label{eq:sugraact} I_{sugra} = {1 \over 16\pi G} \int d^3 x
\sqrt{g} \left(\CR(g) + {2 \over R^2} + \ldots \right) \ . \ee Here,
$R$ is the AdS length scale  and the ellipses denote contributions
of other fields in the $\CN=2$ supergravity multiplet. Specifically,
apart from the metric, these fields include real spin-${3 \over 2}$
gravitino fields, $\psi_L^i$ and $\psi_R^i$, $i=1,2$ as well as two
abelian gauge fields, $a_L$ and $a_R$. In general, if we were
interested in $\CN=(p,q)$ supergravity theory, the corresponding
gauge group would be $SO(p) \times SO(q)$. Thus, in the present
context of $\CN=(2,2)$ theory we have $SO(2) \times SO(2)$ gauge
fields.

In fact, by enlarging the gauge group one can write the entire
supergravity action \eqref{eq:sugraact} as the Chern-Simons action
\cite{Achucarro:1987vz,Achucarro:1989gm}:
\be
\begin{split}
I_{CS} = & {k_L \over 4\pi} \int \tr \left(
A_L \wedge d A_L + {2 \over 3} A_L \wedge A_L \wedge A_L \right) \\
& - {k_R \over 4\pi} \int \tr \left(
A_R \wedge d A_R + {2 \over 3} A_R \wedge A_R \wedge A_R \right) \\
\end{split}
\ee where the gauge fields $A_L$ and $A_R$ take values in the Lie
algebra of the supergroup \be \label{ggroup} G = G_L \times G_R =
OSp(2 \vert 2)_L \times OSp(2 \vert 2)_R\ . \ee Since the bosonic
part of the supergroup $OSp(2 \vert 2)$ is $SO(2) \times
SL(2,\mathbb{R})$, the gauge group (\ref{ggroup}) contains the
classical symmetry\footnote{This symmetry group is the gauge group
of the analogous formulation of $\CN=0$ gravity theory.} group,
$SL(2,\mathbb{R})_L \times SL(2,\mathbb{R})_R$, of the
three-di\-men\-sio\-nal AdS space. In the simple case $k_L = k_R$,
which will be of interest to us in the present paper, one finds the
following relation between the parameters: \be \label{kklr} k_L =
k_R = \frac{R}{16G} \ . \ee Combining this with the Brown-Henneaux
formula $c_L = c_R = \frac{3R}{2G}$ and using our expression for the
central charge $c_L = c_R = 6m$, we can conveniently write
(\ref{kklr}) as \be \label{kmrelation} k_L = k_R = {m \over 4} \ .
\ee Since we take $m$ to be integer, it follows that $k_L$ and $k_R$
take values in $\frac{1}{4} \mathbb{Z}$. This is consistent with the
fact that the bosonic part of our supergroup $OSp(2 \vert 2)$
contains $SL(2,\mathbb{R})$, which is a double cover of the identity
component of $SO(2,1)$; see section 2.1 of \cite{Witten:2007kt} for
further details on the allowed values of $k_L$ and $k_R$.

The equivalence of $\CN=(2,2)$ supergravity and Chern-Simons theory
based on the supergroup (\ref{ggroup}) is valid not only
classically, but to all orders in perturbation theory, as long as
the perturbative expansion starts with a non-degenerate classical
solution. This way of formulating perturbative $\CN=(2,2)$
supergravity will be useful to us in what follows, in particular, in
section \ref{sec:Discussion} where we discuss quantum corrections.

The $\CN=(2,2)$ case is similar to the $\CN=0$ case of Chern-Simons
gravity: There are no local degrees of freedom, but the Chern-Simons
theory does give rise to ``edge states.'' These are $\CN=2$
descendants of the vacuum representation, that is, the irreducible
highest weight representation defined by $(h=0,q=0)$.

The natural generalization of Witten's proposal to $(2,2)$
supergravity in $2+1$ dimensions is that the holographic dual should
be an ``extremal $(2,2)$ superconformal field theory,''  where we
define the latter to be a theory whose partition function is ``as
close as possible'' to the vacuum character of the $\CN=2$ algebra.
The vacuum character of the $\CN=2$ algebra is \cite{Boucher:1986bh}
\begin{equation}\label{eq:vacchar}
\begin{split}
\chi_{vac}^{(m)}(\tau,z) & := {\Tr}_{V_{0,0}} q^{L_0-c/24} e^{2\pi i
(z+\half) J_0} \\
&\,\, = q^{-m/4}(1-q)\prod_{n=1}^\infty \frac{   (1- y q^{n+1/2}) (1-
y^{-1} q^{n+1/2})}{ (1-q^n)^2} \ .\\
\end{split}
\end{equation}
We have shifted $z$ by $1/2$ relative to the standard definition for
later convenience.  The expression in (\ref{eq:vacchar})  is neither
spectral flow invariant, nor modular invariant, and hence more terms
must certainly be added to get a physical partition function.

In \cite{Cvetic} the near horizon geometry of the $D1D5$ system was
investigated and it was observed that the cosmic censorship bound
for the BTZ black hole, which requires $r_\pm \geq 0$ for the two
roots of the lapse function, can be translated into the holographic
conformal field theory   as the bounds
\begin{equation}\label{eq:ccbound}
4m(L_0 - \frac{c}{24}) - J_0^2 \geq 0  \qquad \& \qquad 4m(\tilde
L_0 - \frac{c}{24}) - \tilde J_0^2 \geq 0.
\end{equation}
In \cite{Dijkgraaf:2000fq} the connection of these inequalities to
the conditions on polarity of terms in the partition function was
pointed out. We will assume here that for general   $\CN=(2,2)$
supergravity the cosmic censorship bound continues to be
(\ref{eq:ccbound}). That is,  black hole states must have   $p,
\tilde p \geq 0$, where $p$ and $\tilde p$ refer to the polarity of
the left- and right-moving states ({\it i.e.}, $p =4mn-\ell^2$). In
a theory of ``pure supergravity'' we would certainly want to require
that all states with $p<0$ and $\tilde p<0$  are $\CN=2$ descendents
of the vacuum (or their spectral flow images). These considerations,
then, motivate  our definition of an $\CN =(2,2)$ extremal conformal
field theory to be:

\smallskip

\noindent \textbf{Definition}: An   $\CN=(2,2)$ \emph{extremal
conformal field theory of level $m$} (``$\CN=2$ ECFT'' for short) is
a hypothetical theory  whose partition function is of the form:
\begin{equation}\label{NSsectorNew}
\begin{split}
Z_{NSNS}(\tau,z;\bar \tau, \bar z) & := {\Tr}_{\CH_{NSNS}}
q^{L_0-c/24} e^{2\pi i z J_0}
 \bar q^{\tilde L_0-c/24} e^{2\pi i \bar z \tilde J_0} e^{i \pi (J_0
 - \tilde J_0)} \\
 & \,\, =
\sum_{s, \bar s \in \IZ} SF_s \chi^{(m)}_{\rm vac}(\tau,z)
SF_{\bar{s}} \bar\chi^{(m)}_{\rm vac}(\bar\tau,\bar{z}) \\
&\quad
+ \sum_{s \in \IZ} SF_s \chi^{(m)}_{\rm vac}(\tau,z)
\bar{f}(\bar\tau,\bar z)
+ \sum_{\bar s \in \IZ}
f(\tau,z) SF_{\bar{s}} \bar\chi^{(m)}_{\rm vac}(\bar\tau,\bar{z})
 \\
&\quad +  \sum_{p,\tilde p\geq 0} a(n,\ell;\tilde n , \tilde \ell) q^n
y^{\ell}\bar q^{\tilde n} \bar y^{\tilde \ell} \ .
\end{split}
\end{equation}
Here  the coefficients $a(n,\ell;\tilde n ,\tilde \ell)$ are
integers, and the sum over nonpolar states in the last line means
that \emph{both}  the left and right polarity of the state is
non-negative. The functions $f(\tau,z)$ and
$\bar{f}(\bar\tau,\bar{z})$ describe the contribution of terms
with non-negative polarity with respect to the left and right polarity,
respectively. We need to include such terms since states with either
$p\geq 0$ or $\tilde p \geq 0$ are not polar and are allowed by the
extremality condition.

Using spectral flow (\ref{eq:DefEllipGen}) we can compute
$Z_{RR}(\tau,z;\bar \tau, \bar z)$ for an $\CN=2$ ECFT from
(\ref{NSsectorNew}). The elliptic genus is then obtained upon
setting $\bar{z}=0$. In this limit only those terms   that have
$\bar{q}^0$ contribute. All of these terms have negative polarity,
with the exception of the $\bar{q}^0 \bar{y}^0$ term that has
polarity zero. Thus the elliptic genus of an $\CN=2$ ECFT of level
$m$ is of the form
\begin{equation}
 (2(-1)^m + u) \sum_{\theta \in
\IZ+\half} SF_\theta \chi^{(m)}_{vac} + {\rm Nonpolar} \ ,
\end{equation}
where $u$ is the coefficient of the $\bar{q}^0 \bar{y}^0$ term
coming from $\bar{f}(\bar\tau,\bar{z})$. The factor $2(-1)^m$ is the
limit $\bar z \rightarrow 0$ of the first term in
(\ref{NSsectorNew}), as we will see momentarily.
Using  (\ref{eq:sumc0}) below one can determine the constant to be
$u = 12m-2$. For convenience we drop the overall constant factor
from the right-movers and define:
\be \label{eq:ExtEG}\chi^{(m)}_{\rm ext}(\tau,z) :=\sum_{\theta \in
\IZ+\half} SF_\theta \chi^{(m)}_{vac} + {\rm Nonpolar}\ . \ee

We will call a weak Jacobi form that satisfies (\ref{eq:ExtEG}) an
\emph{extremal elliptic genus}. Because the only unknown terms in
(\ref{eq:ExtEG}) are nonpolar terms we can compute the  polar
polynomial of such an extremal elliptic genus. We will give an
explicit formula for  it in section \ref{subsec:sugrapolarpoly}.
Then, in section \ref{sec:SearchForEG} we investigate whether such a
polar polynomial is consistent with modularity.


\subsection{The extremal polar polynomial}
\label{subsec:sugrapolarpoly}

Let us compute the polar polynomial of a would-be extremal elliptic genus.
We begin by demonstrating the following useful fact:
\begin{equation}\label{eq:ExtPolarPoly}
{\rm Pol} \Bigl( \sum_{\theta \in \IZ+\half} SF_\theta \chi^{(m)}_{vac}\Bigr)
= {\rm Pol}(SF_{1/2}\chi_{vac}^{(m)})\ .
\end{equation}
Indeed, if we apply the spectral flow by $\theta = l + \frac{1}{2}$
to the vacuum character (\ref{eq:vacchar}) we obtain an expression
of the form
\be
\label{polaryy}
(-1)^m\, q^{l(l+1)m} y^{(2l+1)m} (1-q)  \prod_{n=1}^\infty
\frac{ (1- y q^{n+l+1}) (1- y^{-1} q^{n-l})}{(1-q^n)^2}\ .
\ee
We wish to show that this expression contains no polar terms
in the fundamental domain (\ref{eq:polarregion}) for $l \ne 0$.
Without loss of generality, we can assume $l > 0$.
Note that it is not true that (\ref{polaryy}) has no polar terms.
In fact, already the first term $q^{l(l+1)m} y^{(2l+1)m}$
is polar for every $l$; it has polarity $p = - m^2$.
However, it does not belong the polar region $\CP^{(m)}$
since the power of $y$ is not in the allowed range $1\leq \ell \leq m$.

On the other hand, there are terms in (\ref{polaryy})
with $1\leq \ell \leq m$ but, as we show momentarily, these terms are
not polar.
We can simplify the problem a little bit and omit the denominator
in (\ref{polaryy}) and the factor $(1-q)$ which can only increase the polarity.
Then, our goal is to show that
\be
\label{polarzz}
q^{l(l+1)m} y^{(2l+1)m}  \prod_{n=1}^\infty
(1- y q^{n+l+1}) (1- y^{-1} q^{n-l})
\ee
has no polar terms in the range $1\leq \ell \leq m$. From the above
discussion, we already know that the term $q^{l(l+1)m} y^{(2l+1)m}$
is polar. We can combine it with the terms from factors $(1- y q^{n+l+1})$
and $(1- y^{-1} q^{n-l})$ for various $n$ to bring the power of $y$
to the desired range. Since $l$ is assumed to be positive, it is easy
to see that the terms coming from factors $(1- y q^{n+l+1})$ can be
ignored, while from $\prod_{n=1}^\infty (1- y^{-1} q^{n-l})$
we need to collect at least $2lm$ factors of $y^{-1}$ to bring
the overall power of $y$ to the desired range. The most economical
way to do this (which yields the minimal increase in polarity)
is to collect the factors in the infinite product with the smallest
powers of $q$. These are the terms with $n=1, \ldots, 2lm$:
\be
q^{l(l+1)m} y^{(2l+1)m} \prod_{n=1}^{2lm} y^{-1} q^{n-l}
= q^{(2lm - l + 2)lm} y^{m} \ .
\ee
The resulting term has polarity $p = 4(2lm - l + 2)lm^2-m^2$
which satisfies $p>0$ for any $l,m \ge 1$.
It is easy to see that including other factors from the infinite
product in (\ref{polarzz}) only increases the polarity further.

Having proven (\ref{eq:ExtPolarPoly}) we now define
\begin{equation}\label{eq:defofpm}
p^{(m)}_{\rm ext}:=(-1)^m\,  {\rm Pol}SF_{1/2}\chi_{vac}^{(m)}\ .
\end{equation}
On the other hand, setting $l=0$ in (\ref{polaryy}) one finds
\begin{equation}\label{eq:expsugra}
(-1)^m SF_{1/2}\chi_{vac}^{(m)}= (1-q)y^m   \prod_{n=1}^\infty
\frac{ (1- y q^{n+1}) (1- y^{-1} q^{n})}{  (1-q^n)^2}\ .
\end{equation}
The Fourier expansion of (\ref{eq:expsugra}) begins:
\begin{equation}\label{expSF}
y^m + q (  y^m- y^{m-1} )
 +  q^2 (  - 2 y^{-1 + m} +   3 y^m -  y^{ 1 + m}) +\cdots \ .
\end{equation}

The first few polar polynomials follow easily from  (\ref{expSF})
since the   polar terms for index $m$ have $n\leq
\lfloor\frac{m}{4}\rfloor$. In this way we find that the first few
polar polynomials are:
\begin{eqnarray}
p_{\rm ext}^1 &   = & y\\
p_{\rm ext}^2 &   = & y^2\\
p_{\rm ext}^3 &   = & y^3\\
p_{\rm ext}^4 &   = & y^4\\
p_{\rm ext}^5 &   = & (1+q)y^5\\
p_{\rm ext}^6 &   = & (1+q)y^6-q y^5\\
p_{\rm ext}^7 &   = & (1+q)y^7-q y^6\\
p_{\rm ext}^8 &   = & (1+q)y^8-q y^7\\
p_{\rm ext}^9 &   = & (1+q+3q^2)y^9-q y^8\\
p_{\rm ext}^{10} &   = & (1+q+3q^2)y^{10}-(q +2q^2)y^9\\
p_{\rm ext}^{11} &   = & (1+q+3q^2)y^{11}-(q +2q^2) y^{10}\\
p_{\rm ext}^{12} &   = & (1+q+3q^2)y^{12}-(q +2q^2) y^{11}\ .
\end{eqnarray}
%


\section{Experimental search for the extremal elliptic genus}
\label{sec:SearchForEG}

Since $P(m)> j(m)$ for $m \geq 5$, and since  eq.~(\ref{eq:expsugra})
does not have any obvious modular properties, it
is far from obvious that (\ref{eq:ExtPolarPoly}) is the polar
polynomial of a true weak Jacobi form. In this section we describe
  numerical  results suggesting that in fact, for all
but finitely many $m$ it is not in the image of ${\rm Pol}$ applied
to $\tilde J_{0,m}$.  We will find that there are actually some
``exceptional'' cases where it is in the image for $m\geq 5$. In
section \ref{sec:SearchAnalytic} we will show analytically that
there can only be a finite number of such exceptional cases.   That
might seem to obviate the need for the present section, but the
methods we employ here will prove very useful when we come to
describe nearly extremal theories in section
\ref{sec:NearExtremalCFT}.

Choose a basis $\phi_i$, $i=1,\dots, j(m)$  for $\tilde J_{0,m}$. We
are searching for real numbers $x_i$ such that
\begin{equation}
\sum_{i=1}^{j(m)} x_i {\rm Pol}(\phi_i) = p^{(m)}_{\rm ext}\ .
\end{equation}

A useful way of trying to solve this equation is the following. We
choose a polarity-ordered basis of monomials $q^n y^\ell$ for $V_m$,
that is the basis monomials $q^{n(a)} y^{l(a)}$ where $a=1,\dots,
\dim V_m=P(m)$ so that polarity increases as $a$ increases, and
terms with the same polarity are ordered in increasing powers of
$y$. For example for $a=1$ the most polar term is $y^m$. A
polarity-ordered basis for $V_5$ would be
\begin{equation}
y^5, y^4, y^3,q y^5, y^2, y^1
\end{equation}
with $a=1,\dots, 6$.  The polarity-ordered  basis will be very
useful for our discussion of $\beta$-extremal $\CN=2$ conformal
field theories in section \ref{sec:NearExtremalCFT}.

Having chosen these two bases we can define a matrix $N_{ia}$ of
dimensions $j(m)\times P(m)$ from the expansion
\begin{equation}\label{eq:DefNMatrix}
{\rm Pol}(\phi_i) = \sum_{a=1}^{P(m)} N_{ia} q^{n(a)} y^{\ell(a)}.
\end{equation}
Similarly, we can define   coefficients $d_a$ by
\begin{equation}\label{fullset}
p_{\rm ext}^{(m)} = \sum_{a=1}^{P(m)} d_a q^{n(a)} y^{\ell(a)}.
\end{equation}

Thus, we are  trying to solve the linear equations
\begin{equation}\label{eq:partset}
\sum_{i=1}^{j(m) } x_i N_{ia} = d_a, \qquad\qquad a=1,\dots, P(m).
\end{equation}

It should be stressed that even if we can find a solution $x_i$ to
(\ref{eq:partset}) we are far from establishing the existence of an
$\CN=2$ extremal theory. If a solution exists then the next test we
should apply is to see whether the resulting form $\sum x_i \phi_i$
has \emph{integral} Fourier coefficients. Integrality is clearly a
necessary condition for any candidate elliptic genus since it arises
in conformal field theory from the trace on a Hilbert space.

Using a computer (and the explicit basis (\ref{natbasis}) above)
we have examined equation (\ref{eq:partset}) for  $1\leq m\leq 36$.
We have found that there is a solution $x_i$ in rational numbers
for $1\leq m \leq 5$ and for  $m=7,8,11,13$, but there is no
solution for $m=6,9,10$ and $14 \leq m \leq 36$.\footnote{The
arguments of section 5 demonstrate that there can only be finitely
many solutions. Using the constraints of that section it is easy to
check that there are no further solutions up to $m\leq 400$. This
suggests that the above list is in fact complete.}
Moreover, remarkably, for those values of $m$ which give a solution,
the Fourier coefficients we have explicitly evaluated turn out to be
integral.

The simplest example is the case $m=1$, in which case
$\chi^{(1)}_{ext} = \tilde \phi_{0,1}$.
The next simplest case, $m=2$ yields
\be
\label{mtwoext}
\chi^{(2)}_{ext} = \frac{1}{6} (\tilde\phi_{0,1})^2
+\frac{5}{6} (\tilde\phi_{-2,1})^2 E_4\ .
\ee
Although it is not obvious, one can prove that the Fourier
coefficients are all integral.
Indeed, the claim that this expression has integer Fourier
coefficients is equivalent to the statement
\be
\label{congrxxx}
(\tilde\phi_{0,1})^2 + 5 (\tilde\phi_{-2,1})^2 E_4 = 0 ~~{\rm mod}~6\ .
\ee
In order to prove this, it is convenient to note
(see (\ref{efour}) and (\ref{esix})) that:
$$
E_4=1 ~{\rm mod}~6 \ , \quad E_6=1 ~{\rm mod}~6\ .
$$
Moreover, from (\ref{cuspgen}) it also follows that
$\phi_{10,1} = \phi_{12,1}$ mod 6, which in turn implies
$\tilde \phi_{-2,1} = \tilde \phi_{0,1}$ mod 6, {\it cf.}\
(\ref{phiphitildeexp}).
Substituting this into $(\tilde\phi_{0,1})^2 + 5 (\tilde\phi_{-2,1})^2$
and using the fact that $\tilde\phi_{0,1}$ and $\tilde \phi_{-2,1}$
have integer Fourier coefficients we therefore demonstrate (\ref{congrxxx}).

When we use the basis (\ref{natbasis}) the solutions $x_i$ are rational
numbers with increasingly large denominators as $m$ increases.
For example, already the next case, $m=3$, looks like
\be
\label{mthreeext}
\chi^{(3)}_{ext} = \frac{1}{48}(\tilde\phi_{0,1})^3
+\frac{7}{16} \tilde\phi_{0,1} (\tilde\phi_{-2,1})^2 E_4
+ \frac{13}{24} (\tilde\phi_{-2,1})^3 E_6 \ .
\ee
Even though the coefficients $x_i$ of every monomial
$(\tilde \phi_{-2,1})^a (\tilde \phi_{0,1})^b E_4^c E_6^d$
are rational numbers, the Fourier coefficients $c(n,\ell)$ are integers.
In order to show this, as in the previous example, we express
this as the following statement
\be
\label{congrxxxxx}
(\tilde\phi_{0,1})^3
+ 21 \tilde\phi_{0,1} (\tilde\phi_{-2,1})^2 E_4
+ 26 (\tilde\phi_{-2,1})^3 E_6 = 0 ~~{\rm mod}~48 \ .
\ee
Then, using (\ref{efour}) we note that $E_4 = 1$ mod 48,
so we can ignore $E_4$ in this computation.
It is not true, however, that $E_6 = 1$ mod 48.
Instead, from (\ref{esix}) we find that $E_6^2 = 1$ mod 48.
According to (\ref{phiphitilde}) and (\ref{cuspgen}),
this implies the following identity,
$$
\tilde\phi_{-2,1} = \tilde\phi_{0,1} E_6  ~~{\rm mod}~48\ ,
$$
which, after substituting in the LHS of (\ref{congrxxxxx}),
proves the desired result.


Using the basis of weak Jacobi forms described in section
\ref{sec:ConstructionNearlyExtremal} below one can check that for the
``miraculous'' values $m=5,7,8,11,13$ the solution does indeed have the
property that all the Fourier coefficients $c(n,\ell)$ are integers.


\section{The extremal elliptic genus does not exist for $m$ sufficiently large}
\label{sec:SearchAnalytic}

In this section we give an analytic proof that there is no weak
Jacobi form in $\tilde J_{0,m}$ satisfying (\ref{eq:ExtEG}) for $m$
sufficiently large. Since this section is rather long and technical
let us summarize the main idea here.
Using the spectral flow symmetry one can determine the
NS sector character (without an insertion of $y^{J_0}$ or $(-1)^F$)
from the elliptic genus. This character is a modular form for a
congruence subgroup $\Gamma_\theta$ of the modular group.
It is therefore highly constrained, and as in the case discussed in
\cite{Witten:2007kt}, determined by the coefficients of the negative powers
of $q$, which in turn are fixed by the polar terms of the original elliptic
genus. On the other hand, given the full NS sector character,
we can also determine from it, by a suitable modular transformation,
the R-sector character (without an insertion of $(-1)^F$), and thus,
in particular, its leading term in the $q$-expansion. This coefficient
is however also directly determined by the extremal hypothesis
and a sum rule (\ref{eq:GritsenkoIdent})   for Fourier coefficients.
The two ways of evaluating the same coefficient lead to a non-trivial
constraint on $m$, equation (\ref{eq:constraint}).
Using properties
of modular forms one can show that this constraint is violated for
sufficiently large $m$. The argument must be broken up into cases:
$m$ odd, $m=2\ \mod\ 4$ and $m=0\ \mod\ 4$, of which the last case is
technically the most difficult. In this section we will give the main
line of argument, whereas the technical details can be found in
appendix~\ref{s:estimates}.

\subsection{NS-sector elliptic genus}
\label{subsec:NSsectorEG}

Suppose $\chi(\tau,z)$ is the elliptic genus of a CFT with $\chi \in
\tilde J_{0,m}$. By spectral flow we define the ``NS sector elliptic
genus'' to be\footnote{Note that unlike the NS vacuum character
(\ref{eq:vacchar}), $\chi_{NS}(\tau,z)$ does not involve the shift of $z$ by $1/2$.}
\begin{equation}\label{eq:NSellip}
\chi_{NS}(\tau,z) := e\left[m \left(\frac{\tau}{4} + z+ \half \right) \right]
\chi\left(\tau, z + \frac{\tau}{2} + \half  \right) \ .
\end{equation}
Using the transformation properties of a Jacobi form it follows
easily that
\begin{equation}\label{eq:NSegtmrn}
\begin{split} \chi_{NS}(-1/\tau, z/\tau) & = (-1)^m
e\left(\frac{m z^2}{\tau}\right)\, \chi_{NS}(\tau,z)\\
\chi_{NS}(\tau+2,z) & = (-1)^m\,  \chi_{NS}(\tau,z) \ .\\
\end{split}
\end{equation}
If we put $z=0$ we thus obtain simple transformation laws for
$\chi_{NS}(\tau):= \chi_{NS}(\tau,0)$  under   the congruence
subgroup $\Gamma_\theta = \langle T^2, S\rangle$. (In this section
we consider the modular group to be $PSL(2,\IZ)$.) For $m$ even we
have a strict modular function and for $m$ odd we have a function
with multiplier system given by $-1$ on the two generators.

\EPSFIGURE{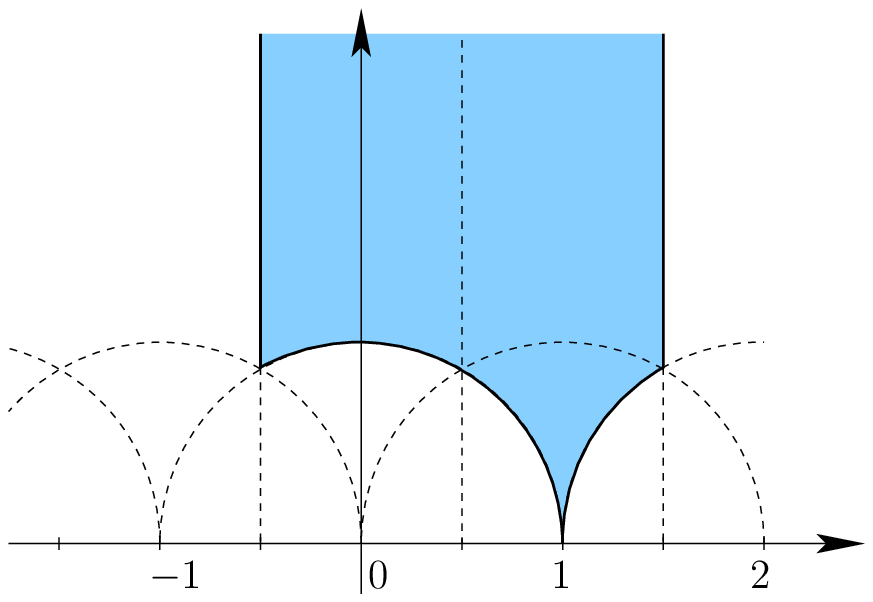,height=6.3cm,angle=0,trim=0 0 0 0}%
{The fundamental domain $\CF_\theta$ of the genus zero subgroup
$\Gamma_\theta$ of $\Gamma$.
\label{domainfig} }


To begin, let us sketch a few mathematical facts.  The group
$\Gamma_\theta$ is a genus zero subgroup of $\Gamma$. It has modular
domain $\CF_\theta = \CF \cup T \cdot \CF \cup TS \cdot \CF$ shown
in figure \ref{domainfig}. Note there are two cusps, equivalent to
$\tau=i \infty$ and $\tau =1$.

Since $\IH/\Gamma_\theta$ is genus zero the function field has a
generator $\hat K(\tau)$ which can be uniquely specified (up to an
additive and multiplicative constant) by demanding that $\hat K$
takes $i \infty $ to $\infty$.\footnote{Such a function for a genus
zero congruence subgroup is often referred to as a ``Hauptmodul.''}
An explicit choice is:
\be \label{eq:KhatHaupt}
\hat K(\tau):=
\frac{\vartheta_3^{12}}{\eta^{12}}(\tau) =
\frac{\Delta^2(\tau)}{\Delta(2\tau)\Delta(\tau/2)} = q^{-1/2} + 24 +
276 q^{1/2} + \cdots \ .\ee
The expansion of $\hat K$ around the cusp at $\tau=1$ is obtained by
writing $\tau = 1- \frac{1}{\tau_r}$ and observing that
\begin{equation}\label{eq:simplektilde}
\hat K(\tau) := \tilde K(\tau_r) = -
\frac{\vartheta_2^{12}}{\eta^{12}}(\tau_r) = - 2^{12} q_r + \cdots \ ,
\end{equation}
where $q_r = e(\tau_r)$.

In order to work with the case of $m$ odd it will be useful to
consider the index two subgroup\footnote{To prove the subgroup is
index two  note that for all $n\in \IZ$,  $T^{4n}$, $S T^{4n+2}$,
$T^{4n+2}S$ and $S T^{4n}S$ are  in $\tilde\Gamma_\theta$. Then use
induction on the length of the word in $S,T^2$. Recall that in this
section modular transformations are regarded as elements of
$PSL(2,\IZ)$. } $\tilde \Gamma_\theta := \langle T^4, S T^2\rangle$
such that $\Gamma_\theta = \tilde \Gamma_\theta \cup S \cdot \tilde
\Gamma_\theta$. This is again a genus zero subgroup, and its
Hauptmodul is the   NS-sector character of $\tilde \phi_{0,1}$ (\ie
the elliptic genus for $K3$ divided by two). Using the definition
(\ref{eq:NSellip}) with $m=1$ and $\chi=\tilde \phi_{0,1}$ and
putting $z=0$ one finds
\begin{equation}
\kappa(\tau):= \left(\frac{2 \vartheta_4}{\vartheta_2}\right)^2 -
\left(\frac{2 \vartheta_2}{\vartheta_4}\right)^2 = q^{-1/4}(1- 20
q^{1/2} + \cdots ) \ .
\end{equation}
This function satisfies $\kappa\vert_S = - \kappa$ and
$\kappa\vert_{T^2}=-\kappa$, and is thus odd under the Deck
transformation $\IH/\tilde\Gamma_\theta \to \IH/\Gamma_\theta$.
Indeed,
\begin{equation}
\kappa^2(\tau) = \hat K(\tau)-64\ ,
\end{equation}
giving the explicit double cover.  Near the Ramond cusp $\kappa$ has
the expansion
\begin{equation}
\kappa(1- 1/\tau_r):= \tilde\kappa(\tau_r) = - 4 i \biggl[
\left(\frac{\vartheta_3}{\vartheta_4}\right)^2 +
\left(\frac{\vartheta_4}{\vartheta_3}\right)^2 \biggr](\tau_r)
= -8i \Bigl( 1+ 32 q_r + \CO(q_r^2)\Bigr) \ .
\end{equation}

Now, $\chi_{NS}$ has no singularities for $\tau\in \IH$, and,
moreover, using again the transformation laws of a Jacobi form
\be \label{eq:expdcusp1} \chi_{NS}(1-1/\tau_r)
=e\left(- \frac{m}{4} \right) \,
\chi(\tau_r, \half)
= e\left(- \frac{m}{4} \right)\,  \sum_{n,\ell} c(n,\ell)\,
(-1)^\ell\,  q_r^n \ .
\ee
By unitarity the sum is over $n\geq 0$ and hence $\chi_{NS}(\tau)$
must be a polynomial in $\kappa(\tau)$. This polynomial will be even
for $m$ even and odd for $m$ odd. Moreover, the polynomial is
fixed by the coefficients of the nonpositive powers of $q$. Those
coefficients in turn are related to the polar contributions to
$\chi(\tau,z)$. To demonstrate the relationship note that
\begin{equation}\label{eq:znssub}
\chi_{NS}(\tau) = \sum_{n,\ell} c(n,\ell)\,
(-1)^{m+\ell}\,
q^{\frac{m}{4} + n + \frac{\ell}{2}} \ .
\end{equation}
Now write:
\begin{equation}\label{eq:simple}
4mn - \ell^2 = 4m \left(\frac{m}{4} + n + \frac{\ell}{2}\right) - (m+\ell)^2 \ .
\end{equation}
The nonpolar terms in $\chi(\tau,z)$  have $4mn-\ell^2\geq 0$
and therefore from (\ref{eq:simple}) contribute only nonnegative
powers of $q$ in (\ref{eq:znssub}). In fact, they always contribute
positive powers with precisely one exception: when $4mn-\ell^2=0$
and $\ell=-m$. In that case $n=m/4$. Note that this cannot happen if
$m\not=0 \, \mod \, 4$ because $n$ is integral.

\subsection{A nontrivial constraint}
\label{subsec:NontrivialConstraint}

 In this subsection
we   assume $m\not=0\, \mod \, 4$. We  return to a discussion of the
case $m=0\, \mod \, 4$  in subsection \ref{subsec:m0mod4} below.

Our conclusion thus far is that   for $m\not=0 \, \mod\, 4$,
$\chi_{NS}(\tau)$ is a modular function for $\tilde\Gamma_\theta$
such that
\begin{equation}
\chi_{NS}(\tau) =   \sum_{\theta\in \IZ} SF_\theta
\chi_{vac}^{(m)}(\tau,z) \vert_{z=\half} + \CO(q^{1/4}) \ .
\end{equation}
One easily finds that only $\theta=0$ can contribute to negative
powers of $q$ and hence we can simplify this equation to
\begin{equation}\label{eq:chivdef}
\chi_{NS}(\tau) =   q^{-m/4}(1-q) \prod_{n=1}^\infty \frac{ (1+
q^{n+1/2})^2} { (1-q^n)^2} + \CO(q^{1/4}) \ .
\end{equation}
This has expansion
\begin{equation}\label{eq:chivdefexp}
   q^{-m/4}\left( 1 + q + 2 q^{3/2} + 3 q^2 + 4 q^{5/2} + 6 q^3 + \cdots \right) \ .
\end{equation}
While the expression on the RHS of (\ref{eq:chivdef}) is not
modular, it can be written as:
\begin{equation}\label{eq:sxpr}
\chi_{NS}(\tau) =   q^{-m/4} \frac{ 1- q^{1/2}}{1+q^{1/2}} q^{1/8}
\frac{\vartheta_3}{\eta^3} + \CO(q^{1/4}) \ .
\end{equation}
Now we can write an explicit formula for $\chi_{NS}(\tau)$. Define
expansion coefficients:
\begin{equation}\label{eq:hatildef}  q^{-m/4} q^{1/8} \frac{ 1- q^{1/2}}{1+q^{1/2}}
  \frac{\vartheta_3}{\eta^3}= \sum_{\alpha=-m/4}^\infty
\tilde h(\alpha) q^{\alpha} \ .
\end{equation}
Note that $\tilde h(\alpha)$ is only nonzero for $\alpha\in \half
\IZ$, for $m$ even and $\frac{1}{4} + \half \IZ$ for $m$ odd.
For $\alpha \in \frac{1}{4}\IZ_+$ let  $\wp_\alpha$ be the unique
polynomial of degree $4\alpha$ such that
\begin{equation}
\wp_\alpha(\kappa) = q^{-\alpha} + \CO(q^{1/4}) \ , \qquad
\alpha \in \frac{1}{4}\IZ_+ \ .
\end{equation}
Then for $m\not=0\, \mod\, 4$
\be \chi_{NS} = \sum_{\alpha=-m/4}^0 \tilde h(\alpha)\,
\wp_{-\alpha}(\kappa)\ . \label{htildeexpd}
\ee
On the other hand, if we expand around the cusp $\tau=1$ then, by
(\ref{eq:expdcusp1})
\begin{equation}
 \sum_{\alpha=-m/4}^0 \tilde h(\alpha) \,
 \wp_{-\alpha}(\tilde\kappa(\tau_r))
 = e^{-i \pi m/2} \sum_{n,\ell\in\IZ} c(n,\ell)\,
(-1)^\ell \, q_r^n \ .
\end{equation}
In particular, if we take $\tau_r\to i \infty$, then we arrive at
the key constraint:
\begin{equation}\label{eq:constraint}
L:= \sum_{\alpha=-m/4}^0 \tilde h(\alpha) \, \wp_{-\alpha}(-8i)
 = e^{-i \pi m/2} \sum_{\ell}   c(0,\ell)\, (-1)^\ell \ .
\end{equation}
The argument for the non-existence of the extremal elliptic genus is
based on showing that, for large $m$, the left-hand side and
right-hand side of (\ref{eq:constraint}) have different growth
rates. As we shall see momentarily, the right-hand side is always an
affine linear function of $m$, while the left-hand side grows
exponentially for  $m=2\, \mod\, 4$; for $m$ odd, the left-hand side
grows also linearly in $m$, but the coefficient is different.
\smallskip

Let us first establish the growth property of the right-hand side.
By the ansatz for pure supergravity we know that the only
nonzero polar coefficients $c(0,\ell)$ occur for $\ell = \pm m$ and
are given by $1$. The coefficient $c(0,0)$ is \emph{not} polar.
Fortunately, Gritsenko has proven a useful identity for the Fourier
coefficients of weak Jacobi forms of index $m$ \cite{gritsenko-991}:
\footnote{The proof is very simple: $\exp[-8\pi^2 m G_2(\tau)z^2]
\chi(\tau,z)$ transforms as a weight zero modular form. Therefore
the coefficients of $z^{2n}$ in the Taylor series around $z=0$
transform like forms of weight $2n$. In particular the coefficient
of $z^2$ must vanish, since there are no modular forms of weight
two.}
\begin{equation}\label{eq:GritsenkoIdent}
m \sum_\ell c(0,\ell) = 6 \sum_\ell \ell^2 c(0,\ell) \ .
\end{equation}
Using (\ref{eq:GritsenkoIdent}) and (\ref{eq:expsugra}) we can solve
for $c(0,0)$ to get $c(0,0) = 12m -2$, and therefore
\begin{equation}\label{eq:sumc0}
 \sum_\ell c(0,\ell)(-1)^\ell = 12m-2 + 2(-1)^m = \begin{cases} 12m
 & m~~ {\rm even } \\  12m-4 & m ~~ {\rm odd.} \\ \end{cases}
 \end{equation}
In particular, the right-hand side of (\ref{eq:constraint}) grows
linearly with $m$.

 Now let us turn to the left-hand side of (\ref{eq:constraint}).
Observe that this is the $q^0$ term in the $q$-expansion of
\begin{equation}
\left( \sum_{\alpha \geq -m/4} \tilde h(\alpha) q^{\alpha} \right)
\left( \sum_{n\geq 0} q^{n/4} \wp_{n/4}(-8i) \right) \ .
\end{equation}
On the other hand, using the fact that $\kappa$ is a Hauptmodul  one
can show,  that \footnote{Write $\wp_{\alpha}(z) = \oint_C
\frac{\wp_\alpha(\ell)}{\ell-z} \frac{d\ell}{2\pi i}$ where the
contour is on a large circle $C$ in the $\ell$ plane. Now make the
change of variables $\ell\to \ell(r) := r^{-1} - 20r + \cdots$  so
that $\ell(q^{1/4}) = \kappa$.   This gives a one-one map of  $C$ to
a small circle $C'$ around the origin. Using $\wp_{\alpha}(\ell(r))
= r^{-4\alpha} + \CO(r)$, and taking the circle to be small we see
$\wp_{\alpha}(z) = - \oint_{C'}\frac{\ell'(r)
r^{-4\alpha}}{\ell(r)-z} \frac{dr}{2\pi i} $. It is now
straightforward to sum the series and apply Cauchy's theorem to
arrive at (\ref{eq:GannonId}). We thank Terry Gannon for pointing
out this crucial identity to us.  }
\be\label{eq:GannonId} \sum_{n=0}^\infty q^{n/4} \wp_{n/4}(z) =
\frac{4 q {d\over dq} \kappa}{z-\kappa}\ . \ee
In order to apply this to our problem we use the identities
\footnote{To prove these identities note that $(24q{d \over
dq}-E_2)\vartheta_2$ must be a weight $5/2$ modular form for
$\Gamma(2)$ and hence is a polynomial of degree $5$ in $\vartheta_2,
\vartheta_3,\vartheta_4$. Moreover, the $q$ expansion has only
coefficients $q^{\frac{1}{8} +n}$ with $n$ integer. Together with
the transformation property under $\tau \to \tau +1$ this fixes it
to be of the form $\vartheta_2 ( a (\vartheta_3^4 + \vartheta_4^4) +
b \vartheta_3 \vartheta_4 (\vartheta_3^2 + \vartheta_4^2)+ c
\vartheta_3^2 \vartheta_4^2)$ for some constants $a,b,c$. Now,
matching the first 3 coefficients of the $q$ expansion on the left
and right hand sides we find $a=1, b=c=0$. The other two equations
now follow by modular transformations.
These identities also have nice interpretations in terms of
massless free fermions on a two-dimensional torus. One can
compute the expectation value of their energy either by differentiating
their partition function or by evaluating the energy-momentum tensor
using the fermion two-point function.
Requiring that these two methods produce the same answer implies
these identities \cite{Eguchi:1986sb}.}
\be \label{eq:logderivtheta} \begin{split} 24q{d \over  dq} \log
\vartheta_4 & = E_2 -
(\vartheta_2^4 + \vartheta_3^4) \\
 24q{d \over  dq} \log
\vartheta_3 & = E_2 + (\vartheta_2^4 - \vartheta_4^4) \\ 24q{d \over
dq} \log \vartheta_2 & = E_2 + (\vartheta_3^4 + \vartheta_4^4) \\
\end{split}
\ee
to compute $4 q{d\over dq} \kappa = - 4
\vartheta_3^8/(\vartheta_2^2\vartheta_4^2)$. Using the ``abstruse
identity''  $\vartheta_3^4 = \vartheta_2^4 + \vartheta_4^4$  it
follows that
\be
 \sum_{n=0}^\infty q^{n/4} \wp_{n/4}(-8i)=  (\vartheta_4^2 - i
 \vartheta_2^2)^2 \ .
\ee
Thus, we need to estimate the large $m$ behavior of
\be \label{eq:CoeffToEstimate} L := \biggl[   q^{-\frac{m}{4}+
\frac{1}{8}} \frac{1-q^{1/2}}{1+q^{1/2}}
\frac{\vartheta_3}{\eta^3}(\vartheta_4^2-i \vartheta_2^2)^2
\biggr]_{q^0}  \ . \ee
We estimate the growth behavior of $L$ in appendix~\ref{s:estimates},
and it turns out to be quite different for even and odd.

For $m$ odd, $e^{i \pi m/2} L$ is positive, and is
bounded below by
\be e^{i \pi m/2} L \geq 4 \pi m - 8\pi \sqrt{m-\frac{5}{2}} - 6\pi
\ .
\ee

Since $4\pi>12$, this will asymptotically (\ie for
$m\geq 2000$) grow more quickly than (\ref{eq:sumc0}). We have
checked that among the first $2000$ terms, the two numbers only agree
for $m=1,3,5,7,11,13,19,31,41$. For
$m=1,3,5,7,11,13$ there exists indeed a sugra elliptic genus, while
for $m=19,31,41$ there does not, as we have verified explicitly.
(Note that the fact that the two
numbers agree does not imply that there must exist a sugra elliptic
genus!)

For $m = 2\ \mod\ 4$, $L$ turns out to grow exponentially, so that
(\ref{eq:constraint}) cannot be satisfied for $m$ large enough. For
details of the calculation, see again appendix~\ref{s:estimates}.

\subsection{A constraint for $m=0\ \mod\ 4\ $}
\label{subsec:m0mod4}

 We now turn to the case $m=0\ \mod\,
4$. As we have pointed out above, in this case non-polar terms
contribute to the constant term of $\chi_{NS}$. We thus need to make
the more general ansatz
\begin{equation}\label{eq:0mod4A}
\chi_{NS}(\tau,z) =    q^{-\frac{m}{4} + \frac{1}{8}}
\frac{1-q}{(1+y q^{1/2})(1+y^{-1}q^{1/2})}
\frac{\vartheta_3(\tau,z)}{\eta^3} + d + \CO(q^{1/2})  \ .
\end{equation}
Instead of (\ref{htildeexpd}) we obtain
\be \label{eq:0mod4} \chi_{NS} = \sum_{\alpha=-m/4}^{0} \tilde
h(\alpha) \wp_{-\alpha}(\kappa) + d\ .
\ee
The argument of section \ref{subsec:NontrivialConstraint} can
then be used to fix the value of $d$:
\begin{equation}\label{eq:0mod4B}
d= 12m - \biggl[   q^{-\frac{m}{4}+
\frac{1}{8}}\frac{1-q^{1/2}}{1+q^{1/2}}
\frac{\vartheta_3(\tau)}{\eta^3}(\vartheta_4^4- \vartheta_2^4)
\biggr]_{q^0}\ .
\end{equation}
We obtain an additional constraint on the theory in the following way:
 Let
\begin{equation}
\hat D := \left(y \frac{d}{dy}\right)^2 - \frac{m}{6}E_2\ .
\end{equation}
Then $\hat \chi_{NS}(\tau):=\hat D(\chi_{NS}(\tau,z))\vert_{z=0} $
is a weight two weakly holomorphic modular form for $\Gamma_\theta$
which moreover satisfies
\begin{equation}\label{eq:HKDerR}
\hat\chi_{NS}(1-1/\tau_r) = \tau_r^2 \hat D
(\chi(\tau,z))\vert_{z=1/2}\ .
\end{equation}
The $q_r\to 0$ limit of the coefficient of $\tau_r^2$ of the
right-hand-side of (\ref{eq:HKDerR}) is
\begin{equation}
\sum_{\ell} c(0,\ell)(-1)^\ell \ell^2 - \frac{m}{6}\sum
c(0,\ell)(-1)^\ell = 2m^2 - \frac{m}{6} 12m =0 \ .
\end{equation}
On the other hand, weakly holomorphic modular forms of weight two
for $\Gamma_\theta$ are of the form
\begin{equation}
(\vartheta_2^4-\vartheta_4^4) \times L(\hat K)\ ,
\end{equation}
where $L(\hat K)$ is a Laurent series in $\hat K$. By examining the
Ramond cusp we see that $L(\hat K)$ must be a polynomial in $\hat
K$. Define polynomials $P_a(\hat K) = q^{-a/2} + \CO(q^{1/2})$ for
$a\geq 0$ and
\begin{equation}
\tilde P_a(\hat K) (\vartheta_2^4 - \vartheta_4^4) = \begin{cases} 1
+ \CO(q^{1/2}) & a=0 \\
a q^{-a/2} + \CO(q^{1/2}) & a>0 \ . \\ \end{cases}
\end{equation}
Using (\ref{eq:logderivtheta}) we find
\begin{equation}
2 q{d\over dq} \hat K = \hat K(\vartheta_2^4 - \vartheta_4^4)\ ,
\end{equation}
from which we deduce
\begin{equation}
\tilde P_a(z) = \begin{cases} -1 & a=0 \\ - z P_a'(z) & a >0 \ . \\
\end{cases}
\end{equation}

Define expansion coefficients
\begin{equation}
\hat \chi_{NS}(\tau) = \sum_{\alpha=-m/4} (-2\alpha) x(\alpha)
q^{\alpha} + X(0) \ .
\end{equation}
If the extremal elliptic genus exists then
\begin{equation}
\hat \chi_{NS}(\tau) = \sum_{\alpha<0 }   x(\alpha) \tilde
P_{-\alpha} (\hat K)(\vartheta_2^4 - \vartheta_4^4)  -
X(0)(\vartheta_2^4 - \vartheta_4^4) \ .
\end{equation}
Evaluating at the Ramond cusp we have
\begin{equation}  \tau_r^2\left(X(0)(\vartheta_4^4 + \vartheta_3^4)
- \sum_{\alpha<0} x(\alpha) \tilde P_{-\alpha}(\tilde
K)(\vartheta_4^4 + \vartheta_3^4)\right) \ ,
\end{equation}
and evaluating at $q_r \to 0$ the coefficient of $\tau_r^2$ becomes
simply $2X(0)$ since $\tilde P_\alpha(0)=0$ for $\alpha>0$.
Therefore, $X(0)=0$.

On the other hand, we can deduce the coefficient $X(0)$ directly from
the $q^0$ term of $\hat D \chi_{NS}$.
Expressing $\chi_{NS}$ by (\ref{eq:0mod4A}) and (\ref{eq:0mod4B}) and
then using
\begin{eqnarray}
(y \partial_y)^2\frac{1}{(1+y
q^{1/2})(1+y^{-1}q^{1/2})}\Big |_{y=1} &=& - \frac{2
q^{1/2}}{(1+q^{1/2})^4}\ , \\
y \partial_y \vartheta_3 \Big |_{y=1} &=& 0 \ , \\
(y \partial_y)^2 \vartheta_3 \Big |_{y=1} & =&  2 q \partial_q
\vartheta_3 \ ,
\end{eqnarray}
and (\ref{eq:logderivtheta}), we obtain the constraint
\begin{eqnarray}\label{eq:R1R2constraint}
0 & = &
\left[\hat D \chi_{NS}\right]_{q^0} = \left[(y\partial_y)^2 \chi_{NS} -
\frac{m}{6}E_2\, \chi_{NS}\right]_{q^0} \nonumber \\
& = & - 2 m^2 + \left[ q^{-m/4+1/8} \frac{1-q^{1/2}}{1+q^{1/2}}
\frac{-2q^{1/2}}{(1+q^{1/2})^2}\frac{\vartheta_3}{\eta^3}\right]_{q^0}
\nonumber \\
& & \qquad
-(4m-2) \left[ q^{-m/4+1/8} \frac{1-q^{1/2}}{1+q^{1/2}}
\frac{q \partial_q \vartheta_3}{\eta^3}\right]_{q^0} \nonumber \\
& = & - 2 m^2 - R_1 - (4m-2)\, R_2 \ , \label{0mod4main}
\end{eqnarray}
where $R_1$ and $R_2$ are defined as
\begin{eqnarray}
R_1 &=& \left[ 2\, q^{1/2} \, \frac{(1-q^{1/2})^4}{(1-q)^3}
\frac{
  \vartheta_3}{\eta^3}\right]_{q^{\frac{m}{4}-\frac{1}{8} }}
\\
R_2 &=&  \left[ \frac{(1-q^{1/2})^2}{1-q} \frac{q \partial_q
\vartheta_3}{\eta^3}\right]_{ q^{\frac{m}{4}-\frac{1}{8} }} \ .
\end{eqnarray}
In appendix~\ref{s:estimates} we show that for large enough $m$ both
$R_1$ and $R_2$ are positive.
It is then clear that (\ref{0mod4main}) cannot be satisfied.

\subsection{What are the exceptional values of $m$? }

The results of the previous subsections establish rigorously that
there are at most a finite number of values of $m$ for which a
candidate extremal elliptic genus can exist. The results of section
\ref{sec:SearchForEG} suggest that there are in fact precisely $9$
such values namely $1\leq m \leq 5$, and   $m=7,8, 11,13$. Although
we do not have a rigorous proof, we strongly believe this list to be
complete.

\EPSFIGURE{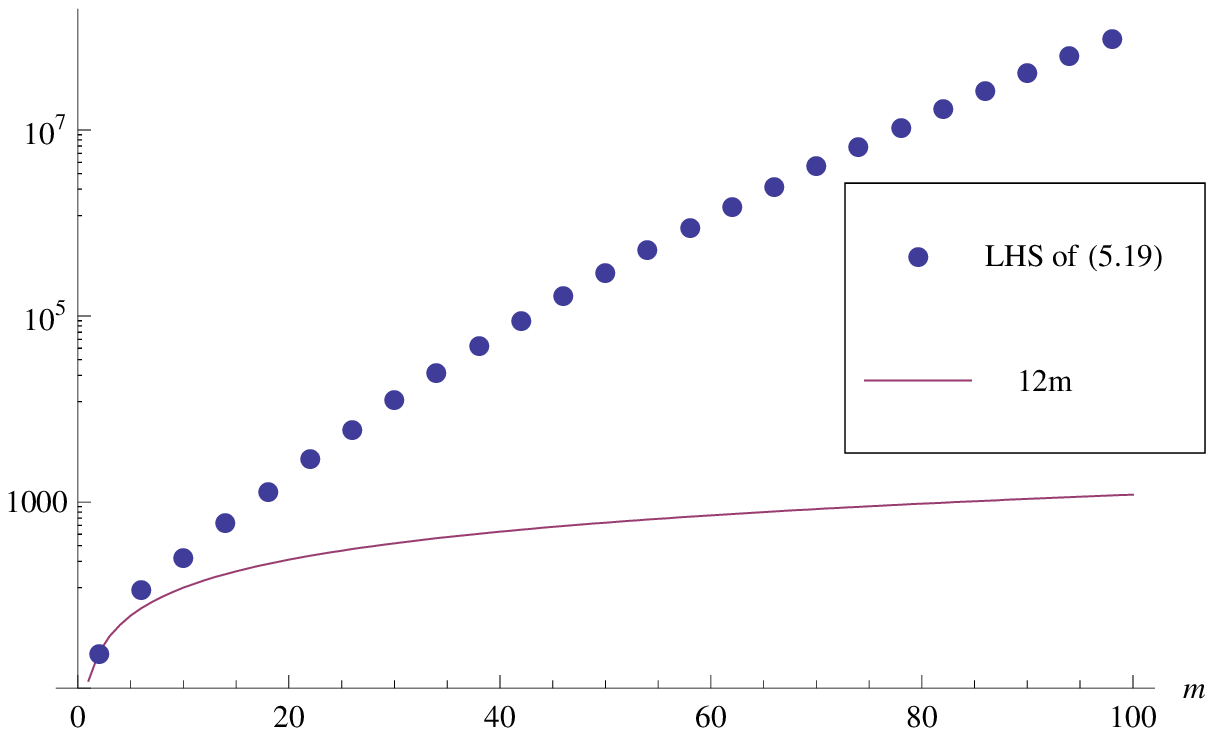,height=6cm,angle=0,trim=0 0 0 0}%
{A semi-log plot of the left-hand side and right-hand side of the
constraint (\ref{eq:constraint}) for $m=2\mod 4$. The left-side
grows exponentially with $m$, as shown on the log plot.
\label{m2mod4} }

\EPSFIGURE{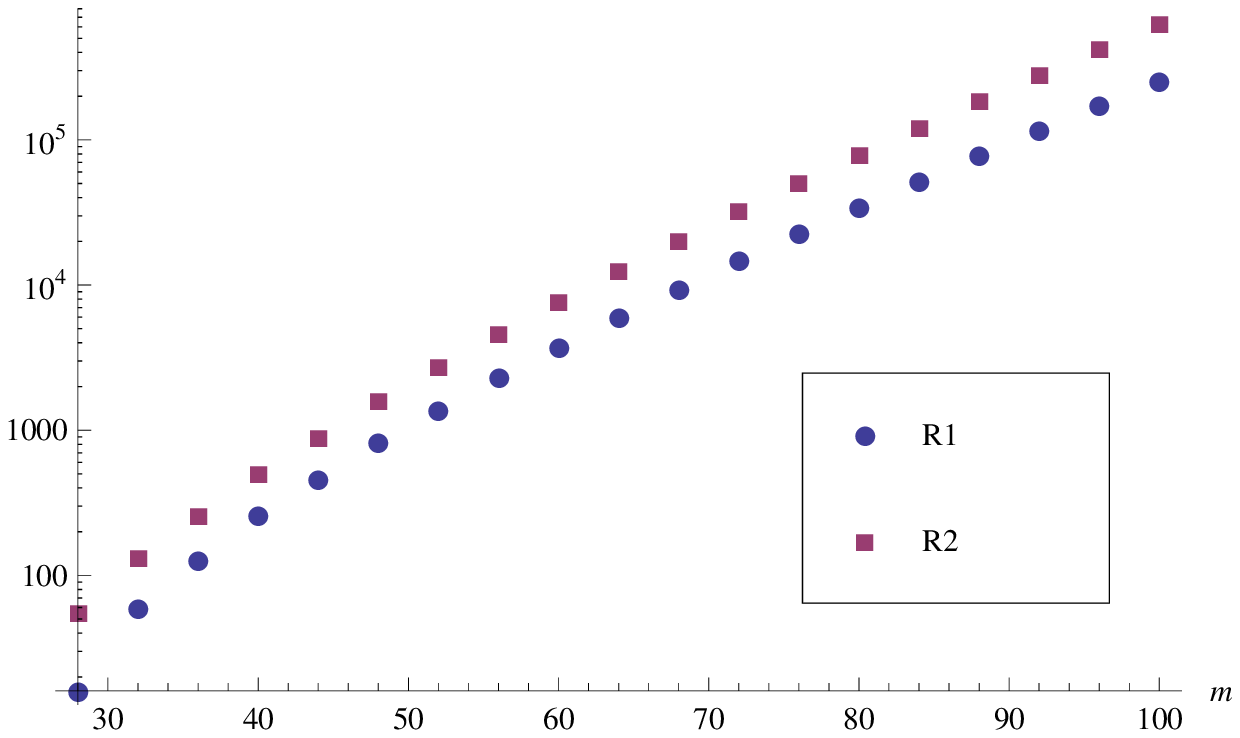,height=6cm,angle=0,trim=0 0 0 0}%
{This figure shows a semi-log plot of  the quantities $R_1$ and
$R_2$ appearing in the constraint (\ref{eq:R1R2constraint}). The
constraint is violated when they are positive. The
plot shows that starting with $m=28$ they are indeed positive
and even exponentially growing. \label{m0mod4} }

As we have mentioned, for $m$ odd we have studied the first 2000
terms and the only possibilities are the values mentioned above. For
$m \sim 2000$ we are well within the regime for which our asymptotic
bounds are valid. For $m$ even we have also examined the constraints
numerically and it appears that $m\geq 36$ is well within the range
of validity of our bounds. See figures \ref{m2mod4} and \ref{m0mod4}
above.


\section{Near-extremal $\CN=2$ conformal field theories}
\label{sec:NearExtremalCFT}

In section \ref{sec:SearchAnalytic} we showed that $\CN=2$ ECFT's,
as we have defined them, at best exist only for a finite number of
exceptional values of $m$. One might object that our definition is
too narrow, and that we should simply modify the definition of an
extremal theory.

In this section we consider one way of modifying the notion of an
extremal theory, by demanding only that some ``significant''
fraction of the polar degeneracies $c(n,\ell)$ coincide with those
predicted from the vacuum character.

Returning to the system of equations (\ref{eq:partset}),
for fixed $m$ define  $k(m)$ to be the largest integer such that
\begin{equation}\label{partset}
\sum_{i=1}^{j(m) } x_i N_{ia} = d_a\ , \qquad\qquad a=1,\dots, k(m)
\end{equation}
admits a solution $x_i$ for which the elliptic genus
$\sum x_i \phi_i$ has an integral Fourier expansion. We would like to
show that we can choose $k(m)$ to be ``close'' to $P(m)$.

Turning again to a numerical analysis, we studied the truncation of
 (\ref{partset}) to the first $j(m)$ equations: $1\leq a \leq j(m)$
 where we ordered the polar terms via their polarity.
 We found that in all cases   $1\leq m\leq 36$ there is indeed a solution
 $x_i$ in rational numbers. Moreover, for all $m$  \emph{except}
  $m=17$ the Fourier expansion coefficients are integral --- in so far
  as we have tested them.  This indicates that
  $k(m)=j(m) + {\cal O}(1)$.\footnote{Note that at least for the
  exceptional
  solutions $m=7,8,11,13$ we have $k(m)>j(m)$.}
  We conjecture that this is the case in general, and in  section
  \ref{sec:modconstspec}, assuming this
  conjecture to be true, we derive an interesting constraint on the
  spectrum of $\CN=2$ CFTs.

For the analysis in section \ref{sec:modconstspec} it
 turns out to be more convenient to define  a ``$\beta$-extremal
$\CN=2$ CFT'' by imposing the less restrictive condition of only
requiring that polar degeneracies are predicted from the vacuum
character in the $\beta$-truncated polar region:
\begin{equation}
\CP_\beta := \{(\ell,n): 1\leq \ell\leq m, n\geq 0,
4mn-\ell^2\leq -\beta \} \ .
\end{equation}
We know that for suitable $\beta$ candidate elliptic genera exist. For
example, if we take $\beta=m^2$ then we can always construct a candidate
elliptic genus. We get a better approximation to an extremal theory by
lowering the value of $\beta$. Therefore, let $P_\beta(m)$ be the
number of independent polar monomials of polarity $\leq -\beta$, and
let  $\beta_*$ be the \emph{smallest } integer $\beta$ such that
\begin{equation}\label{truncated}
\sum_{i=1}^{j(m) } x_i N_{ia} = d_a\ , \qquad\qquad a=1,\dots,
P_\beta(m)
\end{equation}
admits a solution $x_i$ for which $\sum x_i \phi_i$ has
\emph{integral} coefficients in its Fourier expansion. According
to our conjecture $P_{\beta_*}(m) \cong j(m)$. We would therefore
like to estimate the value of $\beta$ for which
$P_\beta(m) = j(m) + \CO(m^{1/2})$ for large $m$.
The computation follows closely the analysis of
section \ref{subsec:CountingPolar}.

We now have
\begin{equation}
P_\beta(m) = \sum_{r=r_0}^m \lceil \frac{r^2-\beta}{4m} \rceil \ ,
\end{equation}
where $r_0 :=\lceil \sqrt{\beta} \rceil$.  As before, we write this
as a sum of three terms,
\be \begin{split}P_\beta(m) = \sum_{r=r_0}^m \frac{r^2-\beta}{4m} -
\sum_{r=r_0}^m ((\frac{r^2-\beta}{4m} ))   + \half \sum_{r=r_0}^m
\left( \lceil \frac{r^2-\beta}{4m} \rceil - \lfloor
\frac{r^2-\beta}{4m} \rfloor\right)\end{split} \ .
\ee
  The first term is
\be \sum_{r=r_0}^m \frac{r^2-\beta}{4m} = \frac{m^2}{12} +
\frac{m}{8} + \frac{1}{24} - \frac{r_0(2r_0-1)(r_0-1)}{24m}- \beta
\frac{(m-r_0+1)}{4m}\ .
\ee
Denote the number of integers $r$ such that $r_0\leq r \leq m$ with
$r^2=\beta\ \mod\ 4m$ by $\nu(m,\beta)$. Unlike the case $\beta=0$ we
cannot write down an exact formula, but it is clear that
asymptotically $\nu(m,\beta)\sim m^{1/2}$.
The second term is
\begin{equation}
  \sum_{r=r_0}^m \lceil \frac{r^2-\beta}{4m} \rceil - \sum_{r=r_0}^m
  \lfloor \frac{r^2-\beta}{4m}
  \rfloor=  m+1-r_0 - \nu(m,\beta)\ .
\end{equation}
For the third term we again use the argument that the  numbers
$((\frac{r^2-\beta}{4m} ))$ are randomly distributed. We thus have a
random walk between $-1/2$ and $+1/2$ and the sum is expected to be
of order $m^{1/2}$.

To conclude, note that  for $\beta = \alpha m$ with $\alpha$ a
constant $0< \alpha < 1$ we have  $r_0\sim m^{1/2}$, so the large
$m$ asymptotics are
\begin{equation}\label{eq:Pbest}
P_\beta(m) = \frac{m^2}{12} + \left(\frac{5}{8}
-\frac{\alpha}{4}\right)m + \CO(m^{1/2})\ .
\end{equation}
Comparing to (\ref{dim_jm}) we see that for large $m$ the reduction
of polarity to obtain the truncated supergravity elliptic genus is
given by $\beta = \half m + \CO(m^{1/2})$.

As in equation (\ref{eq:Pmjm}) above the symbol $\CO(m^{1/2})$ is to
be understood heuristically. It would be worthwhile being more
rigorous about this point.


\subsection{A  constraint on the spectrum of $\CN=2$ theories with
  integral $U(1)$ charges}
\label{sec:modconstspec}

In the previous sections we have found strong evidence that we must
have $P_{\beta_*}(m) \cong j(m)$, and hence by (\ref{eq:Pbest})
\begin{equation}\label{eq:lowerboundbetastar}
\beta_* \geq \frac{m}{2} + \CO(m^{1/2})
\end{equation}
for large $m$.

Now a monomial $q^n y^\ell$ of polarity $\beta$ corresponds by
spectral flow to a state in the NS sector that contributes as
$q^{h-\frac{m}{4}} y^{\ell}$ with
\begin{equation}\label{eq:sfwithbeta}
h = \frac{m}{4} + \frac{ \ell^2}{4m} - \frac{\beta}{4m} \ .
\end{equation}
Therefore, if we accept (\ref{eq:lowerboundbetastar}) then we can
obtain an interesting constraint on the spectrum of a $(2,2)$ $AdS_3$
supergravity with a holographically dual CFT: It must contain at
least one state which is a left-moving $\CN=2$ primary (not
necessarily chiral primary) tensored with a right-moving chiral
primary such that
\be \label{bound}
h < \frac{m}{4} + \frac{\ell^2}{4m}
-\frac{1}{8} +  \CO(m^{-1/2})\ . \ee
It would be interesting and useful to sharpen this bound. However,
we will show in section \ref{sec:ConstructionNearlyExtremal} below
that it \emph{is} possible to construct elliptic genera, which,
after spectral flow, do match the spectrum of the vacuum character
for all conformal weights with $h\leq \frac{ m}{4}$.   There
is no contradiction between this result and (\ref{bound}) because
under $1/2$ unit of spectral flow $0 \leq \vert \ell\vert \leq 2m$
and hence $\frac{\ell^2}{4m}$ could be as large as $m$,
and thus the bound can be as large as $\frac{5m}{4} - \frac{1}{8} +
\CO(m^{-1/2})$.


\section{Construction of nearly extremal elliptic genera}
\label{sec:ConstructionNearlyExtremal}

In this section we consider an alternative basis for the weak Jacobi
forms which has a ``triangular'' nature, allowing us to replace the
polar region $\CP^{(m)}$ by an alternative region $S$. We will see
that for large $m$, $S$ ``approximates'' $\CP^{(m)}$. In the next
section we discuss the possible physical significance of this fact.

It is shown in \cite{gritsenko-991} that there is an \emph{integral} basis of
the ring of weak Jacobi forms of weight zero with integral coefficients
\begin{equation}\label{eq:GritGenerators}
\tilde J^{\IZ}_{0,*} = \IZ[\phi_{0,1}, \phi_{0,2} , \phi_{0,3},
\phi_{0,4} ]/I \ ,
\end{equation}
where $I$ is the ideal generated by the relation
\begin{equation}\label{eq:OneThreeRel}
\phi_{0,1}\phi_{0,3} = 4 \phi_{0,4} + \phi_{0,2}^2\ .
\end{equation}
The generators are elliptic genera of Calabi-Yau manifolds, and
explicit formulae are given in  \cite{gritsenko-991}. In the basis
(\ref{natbasis}) they can be expressed as\footnote{We have redefined
$\phi_{0,4}$ in \cite{gritsenko-991} by a factor of $-1$. }
\begin{eqnarray}
\phi_{0,1} &=& \tilde \phi_{0,1}\\
\phi_{0,2} &=& \frac{1}{24} \tilde \phi_{0,1}^2 - \frac{1}{24} \tilde
   \phi_{-2,1}^2 E_4 \\
\phi_{0,3} &=& \frac{1}{432} \tilde \phi_{0,1}^3 - \frac{1}{144}
\tilde \phi_{0,1} \tilde \phi_{-2,1}^2 E_4 + \frac{1}{216} \tilde
\phi_{-2,1}^3 E_6 \\
\phi_{0,4} &=& \frac{1}{6912} \tilde \phi_{0,1}^4 -
\frac{1}{1152} \tilde \phi_{0,1}^2 \tilde \phi_{-2,1}^2 E_4 +
     \frac{1}{864} \tilde \phi_{0,1} \tilde \phi_{-2,1}^3 E_6
- \frac{1}{2304} \tilde \phi_{-2,1}^4 E_4^2 \ .
\end{eqnarray}

To make the triangular nature of this basis manifest it is useful to
consider the NS sector images of the generators,
\begin{equation}
\hat \phi_{0,m} = (-1)^m q^{m/4} y^m \phi_{0,m}(\tau, z +
\frac{\tau}{2} + \frac{1}{2})\ .
\end{equation}
We now consider ordering the $q,y$ expansion by the leading power of
$q$ and, for each power of $q$ by the \emph{largest} positive power
of $y$. (Recall that $\chi_{NS}(\tau,z)$ is an even function of $z$,
so the positive powers of $y$ determine the negative powers of $y$.)
With this ordering of terms we have
\begin{equation}
\begin{split}
\hat \phi_{0,1} & = q^{-1/4} + \CO(q^{1/4} )     \\
\hat \phi_{0,2} & =  (y+y^{-1})  + \CO(q^{1/2})  \\
\hat \phi_{0,3} & = q^{1/4}(y- y^{-1})^2 + \CO(q^{3/4}) \\
\hat \phi_{0,4} & = 1 + \CO(q^{1/2}) \ . \\
\end{split}
\end{equation}
By (\ref{eq:GritGenerators}) an overcomplete linear basis of $\tilde
J_{0,m}$ is given by
\begin{equation}\label{eq:generalproduct}
(\hat \phi_{0,1})^i(\hat \phi_{0,2})^j (\hat \phi_{0,3})^k (\hat
\phi_{0,4})^l
\end{equation}
with $i+2j + 3k + 4l = m$, $i,j,k,l\geq 0$. In order to obtain a set
of linearly independent basis vectors we distinguish the monomials
in (\ref{eq:generalproduct}) according to  whether $i>k$ or $i\leq
k$ and then use identity   (\ref{eq:OneThreeRel}) to eliminate
$\hat \phi_{0,3}$ or $\hat \phi_{0,1}$, respectively. The result is
that there exists  a vector space basis for $\tilde J_{0,m}$ which
is a disjoint union of two sets $A\amalg B$ with
\begin{equation}\label{eq:Abasis}
A := \{ (\hat \phi_{0,1})^i(\hat \phi_{0,2})^j   (\hat
\phi_{0,4})^l\vert  \quad i>0, j\geq 0, l\geq 0 \qquad i + 2j + 4l
=m\}\ ,
\end{equation}
\begin{equation}\label{eq:Bbasis}
B := \{  (\hat \phi_{0,2})^j  (\hat \phi_{0,3})^k   (\hat
\phi_{0,4})^l\vert  \quad   j\geq 0,k\geq 0 ,  l\geq 0 \qquad 2j +
3k + 4l =m\}\ .
\end{equation}
A tedious but elementary counting argument shows that
\begin{equation}
\begin{split}
 \vert A\vert & =
\begin{cases} \frac{m^2}{16} + \frac{3m}{8} -
\frac{s^2}{16} + \frac{s}{8} + \frac{1}{2} &
m= s\ \mod\ 4, s=1,3 \\
\frac{m^2}{16} + \frac{ m}{4} - \frac{s^2}{16} + \frac{s}{4}
&  m= s\ \mod\ 4, s=0,2 \\ \end{cases} \\
\end{split}
\end{equation}
and $\vert A \vert +\vert B \vert    = j(m)$.


\EPSFIGURE{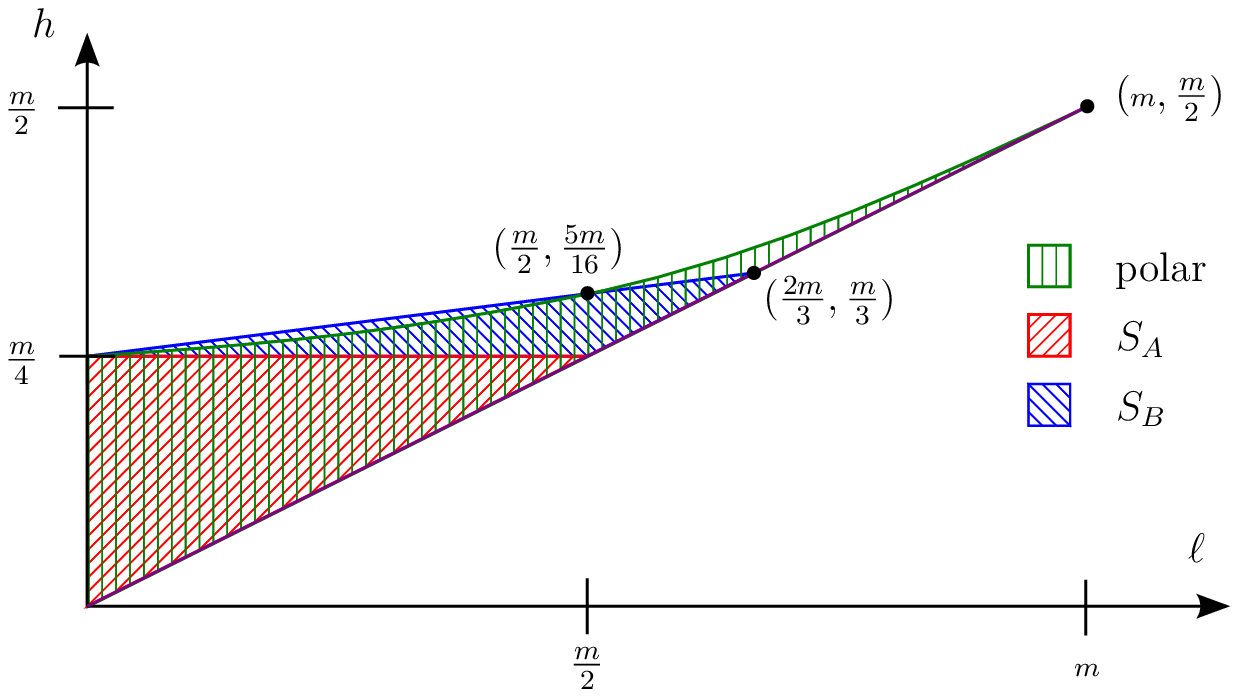,height=8cm,angle=0,trim=0 0 0 0}%
{A comparison of the polar region $\CP^{(m)}$ and the region $S$. The
NS sector polar region is bounded by $\ell\geq 0, h\geq \ell/2,
h\leq \frac{m}{4} + \frac{\ell^2}{4m}$. The region $S$ is the
triangular region, $\ell\geq 0, h\geq \frac{\ell}{2}, h-\frac{m}{4} \leq
\frac{\ell}{8}$,  which itself is a union of two triangular regions
$S_A$ and $S_B$, where $S_A$ is the subregion of $S$ with
$h<\frac{m}{4}$. The polar region contains $S_A$, while  $S_B$ is an
``approximation'' to the remainder.
\label{CompareRegionsfig} }

Now note that the leading expression in the $q,y$ expansion of an
element in the set $A$   is $q^{-i/4} y^{ j} $, while that in the
set $B$ is $q^{k/4} y^{j+2k}$. It thus follows that \emph{an
(NS-sector) Jacobi form of weight zero and index $m$ with integral
Fourier coefficients is uniquely determined by the coefficients of
$q^n y^\ell$ where $(\ell,n)$ run over the set:}
\begin{equation}
S = S_A \amalg S_B \end{equation}
where
\begin{equation}\label{SAdef}
S_A = \Bigl\{ (\ell,n) \vert
  n <0,~~ 0 \leq \ell,~~  n+\frac{m}{4} \geq \frac{\ell}{2}  \Bigr\}
\end{equation}
and
\begin{equation} S_B  =   \Bigl\{ (\ell,n) \vert 0 \leq n ,~~ 8n \leq
\ell,~~
 n+\frac{m}{4} \geq \frac{\ell}{2}  \Bigr\} \ .
\end{equation}
In both $S_A$ and $S_B$ the $(\ell,n)$ are in the lattice $(\ell,n)\in
\IZ \times \frac{1}{4}\IZ$, subject to the quantization condition
\begin{equation}
\left( n+ \frac{m}{4}\right) - \frac{\ell}{2} =0~~ \mod ~ 1 \ .
\end{equation}
(This quantization is equivalent to the statement that in the Ramond
sector the elliptic genus has a Fourier expansion in $q,y$ with
integral powers of $q,y$.)
The regions $S_A$ and $S_B$ in the
$(\ell,h)$ plane are triangles and their union is a triangle.
The full region $S$ can serve as a surrogate for the polar region
$\CP^{(m)}$, as explained in figure \ref{CompareRegionsfig}.

Recall that $n$, the power of $q$ in the NS sector character, is
related to $h$ as $n=h-\frac{m}{4}$. It then follows from
(\ref{SAdef}) that $S_A$ contains all possible points with $h<m/4$
that occur in the NS vacuum character (\ref{eq:vacchar}). Thus it
\emph{is} possible to construct a weak Jacobi form with integral
coefficients whose $q$-expansion agrees with that of an extremal theory
for all NS-sector Virasoro weights up to  $h= m/4$ (for $m$
even) and $h = (m-1)/4$ (for $m$ odd). This fits in very
nicely with the bound (\ref{bound}), which puts an upper bound on
the range of $h$ where all states can be descendants of the
vacuum.


\section{Discussion: quantum corrections to the cosmic censorship bound}

\label{sec:Discussion}

If the pure ${\cal N}=(2,2)$ supergravity is a consistent quantum
theory, its Hilbert space should be spanned by states which can be
identified as excitations of the supergravity fields. One class of
such states are perturbative and normalizable excitations of the
supergravity fields in $AdS_3$, which generate the vacuum
representation in the boundary CFT \cite{Brown:1986nw}. It is
expected that these are the only states up to the cosmic censorship
bound. We define this bound to be the boundary of the region in the
space of energy and charges in  which states corresponding
semiclassically to black hole
solutions can exist. In the classical limit the cosmic censorship
bound is the condition on mass and charges of a black hole such that
there is a regular horizon.

It turns out that the classical cosmic censorship bound is exactly
equal to the upper bound of the polar part of the CFT spectrum
\cite{Dijkgraaf:2000fq}. This was the motivation for the definition
of ${\cal N}=(2,2)$ extremal conformal field theory in section
\ref{secdefinition}. On the other hand, in section
\ref{sec:SearchAnalytic}, we proved that such a conformal field
theory does not exist for sufficiently large $m$. This result,
however, does not immediately rule out the conjectured existence of
pure ${\cal N}=(2,2)$ supergravity   since the cosmic censorship
bound might receive quantum corrections.  That is,
   there might be quantum corrections to the relation between the values
of the mass and charges of those  quantum states whose semiclassical
manifestation are black holes. There are two potential sources for
such corrections, and we will discuss each of them below.

As far as perturbative effects are concerned, the pure supergravity
theory can be treated as the Chern-Simons gauge theory with the
gauge group (\ref{ggroup}). Since the classical equations of motion
of the Chern-Simons theory imply vanishing of the gauge field
strength and since any perturbative corrections to the equations of
motion can be expressed as a polynomial of the field strength and
its covariant derivatives, black hole solutions are not corrected to
all orders in the perturbative (\ie $1/m$)  expansion. However,
values of the mass and charge of a given black hole solution can
receive corrections since computing them requires knowing the action
as well as the equations of motion. In particular, the ``level''
$m$, whose inverse appears in front of the action, can be corrected.
The leading discrepancy between the dimension of the space of polar
polynomials, $P(m)$, and the dimension of the space of weak Jacobi
forms, $j(m)$, \be \label{pjdiff} P(m) - j(m) = {m \over 8} + \CO
(m^{1/2})\ , \ee can be explained if $m$ is shifted by an
appropriate constant by quantum effects. Such a shift is known to
occur in perturbative Chern-Simons gauge theory \cite{WittenJones},
where the level $k$ is shifted at one loop by the dual Coxeter
number of the gauge group, $C_2 (G)$. For the supergroup $OSp(2
\vert 2)$, we have $C_2 = -2$, so that in the present case both
$k_L$ and $k_R$ are shifted as\footnote{One way to think about this
shift is as follows. The supergroup $OSp(2 \vert 2)$ is the
superconformal group of $AdS_2$, and its dual Coxeter number, $C_2$,
can be thought of as the beta-function of the world-sheet
sigma-model defining $AdS_2$ space-time. If instead of $AdS_2$ we
consider a positive curvature space, that is a 2-sphere $S^2$, the
contribution to the beta-function of the world-sheet theory should
have opposite sign and, hence, the opposite shift of $k$. In
particular, for $S^2$, which has the isometry group $SU(2)$, the
shift $k \to k+2$ is familiar in the study of $SU(2)$ Chern-Simons
theory \cite{WittenJones}. In the case of $OSp(2 \vert 2)$
Chern-Simons theory the shift should have opposite sign, therefore
justifying (\ref{kshiftminus}).} \be \label{kshiftminus} k_L \to k_L
- 2 \ . \ee Combining this with equation (\ref{kmrelation}), we can
express this as the shift of $m$, \be m \to m - 8\ , \ee which,
unfortunately, does not account for the difference in
(\ref{pjdiff}). Furthermore, it seems difficult to attribute
sub-leading terms in $(P(m) - j(m))$ to higher order perturbative
effects since sub-leading terms in $P(m)$ contains the arithmetic
function $A(m)$, which does not have a nice $1/m$ expansion (see
footnote 2).

There is another source of corrections which are non-perturbative
in nature. To see this, we note that conformal weights $h$ for
states counted by the elliptic genus are integers, as required by
 modular invariance. This granularity, which is smeared out in
any perturbative analysis, gives rise to an intrinsic ambiguity in
the cosmic censorship bound of $O(1)$ in $h$. Since the boundary of
the polar region in the $(L_0, J_0)$ plane has a length of
order $m$, it is possible that the discrepancy of $P(m)$ and $j(m)$
mentioned in the above is entirely attributed to this granularity.
For example, the bound on $h$ for a new primary state found in
(\ref{bound}) is within $O(1)$ of the cosmic censorship bound.

It is possible that a combination of these two effects resolves the
apparent contradiction between the conjectured existence of pure
${\cal N}=(2,2)$ supergravity and the properties of the elliptic
genus we found in this paper.

Given the close resemblance of the   region $S_A \cup S_B$
identified in section \ref{sec:ConstructionNearlyExtremal} with the
polar region it is natural to ask whether the boundary of that
region might in fact constitute the quantum-corrected cosmic
censorship bound. This seems unlikely to us.  Along the line $h=
\frac{\ell}{8} + \frac{m}{4}$, $0 \leq \ell \leq \frac{2m}{3}$ the
polarity becomes as great as $p=\frac{m^2}{16}$. It seems unlikely
that quantum corrections will modify the mass and charge in such a
way as to change a semi-classical black hole state with such a
polarity to a descendent of the vacuum.

\section{Extremal $\CN=4$ theories}
\label{sec:ExtremeN=4}

The analysis for the case of the pure $\CN=(2,2)$ supergravity
theories is somewhat inconclusive since we cannot rule out that
there are quantum corrections to the classical supergravity ansatz.
The situation is sharper for the case with $\CN=(4,4)$
superconformal symmetry since the possible quantum corrections of
these theories are well constrained \cite{David:2007ak}. Therefore,
in this section  we shall begin to address whether modular
invariance allows for a pure $\CN=(4,4)$ supergravity theory.
Unfortunately, our results are somewhat incomplete.

Following the earlier definition we define an extremal $\CN = (4,4)$
theory to be a theory whose partition function is of the form
 (\ref{NSsectorNew}) where $\chi^{(m)}_{\rm vac}$ is now the vacuum
character of the $\CN=4$ algebra \cite{Eguchi:1987wf,Eguchi:1988af}:
\be \chi^{(m)}_{\rm vac} = q^{-m/4}\, \prod_{n=1}^{\infty}
\frac{(1-yq^{n-1/2})^2 (1-y^{-1} q^{n-1/2})^2}{(1-q^n)} \, \chi(q,y)
\ , \ \ee with
\begin{eqnarray}
\chi(q,y)  & =  &\prod_{n=1}^{\infty} \frac{1}{(1-q^n) (1-y^2 q^n)
  (1-y^{-2} q^{n-1})} \, \nonumber \\
& & \qquad \times \sum_{j\in\mathbb{Z}} q^{(m+1)j^2+j} \,
\Bigl(\frac{y^{2(m+1)j}}{(1-yq^{j+1/2})^2} -
\frac{y^{-2(m+1)j-2}}{(1-y^{-1}q^{j+1/2})^2} \Bigr) \ .
\end{eqnarray}
As in the case of the $\CN=2$ vacuum character, we have evaluated this
expression at $z+\tfrac{1}{2}$. To get rid of the negative powers of
$q$ in the denominator, we can rewrite it as two separate sums over
positive $j$,
\begin{eqnarray}
\chi(q,y)  & = &
\prod_{n=1}^{\infty} \frac{1}{(1-q^n) (1-y^2 q^n)
  (1-y^{-2} q^{n-1})} \, \times \nonumber \\
& & \quad \Bigl[ \sum_{j\geq 0} q^{(m+1)j^2+j} \,
\Bigl(\frac{y^{2(m+1)j}}{(1-yq^{j+1/2})^2} -
\frac{y^{-2(m+1)j-2}}{(1-y^{-1}q^{j+1/2})^2} \Bigr) \nonumber \\
& & \quad \qquad
+ \sum_{j\geq 1} q^{(m+1)j^2+j-1} \,
\Bigl(\frac{y^{-2(m+1)j-2}}{(1-y^{-1}q^{j-1/2})^2} -
\frac{y^{2(m+1)j}}{(1-yq^{j-1/2})^2} \Bigr) \Bigr] \ .
\end{eqnarray}
It is straightforward to read off the polar polynomial from this
expression.

Using the same methods as in section~\ref{sec:SearchForEG}, we have
analyzed whether this polar polynomial can be completed to a weak
Jacobi form. We have performed the analysis for $1\leq m \leq 20$,
and we have found that the only cases where this is possible are
$m=1,2,3,4,5$. (Note that for $1\leq m \leq 4$ this is automatic
since $P(m)=j(m)$.) Thus, apart from a few low level exceptions, we
expect that the pure $\CN=(4,4)$ sugra ansatz is incompatible with
modular invariance. It might  be possible to prove this assertion by
suitably modifying the methods of section \ref{sec:SearchAnalytic},
but the expressions appear to be challenging and we have not
attempted to do so.

An important loophole in our argument is the possibility that there
are zero-modes making the elliptic genus vanish. This might happen when
there is an extension of the chiral algebra and $m$ is odd.  In
order to demonstrate this write the character expansion of the RR
sector partition function as
\begin{equation}
Z_{RR} =  \sum_{1\leq \ell,\tilde\ell\leq m}
c_{\ell\tilde\ell}\chi_{\ell} \overline{\chi_{\tilde \ell}} + c_{00}
\chi_0 \overline{\chi_0}+ \sum_{1\leq \ell\leq m} c_{\ell 0}
\chi_\ell \overline{\chi_0} + \sum_{1\leq \tilde \ell\leq m}
c_{0\tilde\ell} \chi_0 \overline{\chi_{\tilde \ell}} +\cdots
\end{equation}
Here $\chi_\ell$ denote the characters of the unitary massless
representations, with $0\leq \ell \leq m$ denoting twice the spin of
the highest weight vector, and $+\cdots$ refers to terms with a
massive representation on the left or the right. The reason for
separating out the $\ell=0$ spin as special is that its highest
weight vector is not a polar state, whereas the highest weight
vectors of all the other massless representations are polar states.
An extremal theory must have an expansion of the form
\begin{equation}
Z_{RR} =  \chi_{m} \overline{\chi_{\tilde m}} + c_{00} \chi_0
\overline{\chi_0}+ \sum_{1\leq \ell\leq m} c_{\ell 0} \chi_\ell
\overline{\chi_0} + \sum_{1\leq \tilde \ell\leq m}   c_{0\tilde\ell}
\chi_0 \overline{\chi_{\tilde \ell}} +\cdots
\end{equation}
since $\chi_m$ is the spectral flow image of the NS vacuum. Now, the
elliptic genus of $\chi_\ell$ is $(-1)^\ell (\ell+1)$, while that of
the massive representations is zero. Thus, if the elliptic genus
vanishes then, comparing the coefficient of the left-moving vacuum
character $\chi_m$ we see that
\be c_{m0} = (-1)^{m+1}(m+1)\ . \ee
Note that a non-vanishing coefficient $c_{m0}$ implies that the
right-moving chiral algebra is enhanced, as claimed. Also, since
$c_{m0}$ is a positive integer this can only happen when $m$ is odd.
Moreover, by comparing the coefficients of the other left-moving
characters we find the constraints $c_{\ell 0}=0$ for $1\leq \ell
\leq m-1$ and  $\sum_{\tilde\ell=0}^m c_{0\tilde\ell} (-1)^{\tilde
\ell}(\tilde \ell +1) =0$. Since our no-go theorem would apply if
either the   holomorphic or anti-holomorphic elliptic genus is
non-vanishing we might as well assume the anti-holomorphic elliptic
genus also vanishes. In this case we find that $c_{0\ell}=0$ for
$1\leq\ell\leq m-1$ and hence $c_{00}=(m+1)^2$, so that $Z_{RR} =
\vert \chi_m + (m+1)\chi_0\vert^2 + \cdots $. Thus, for extremal
theories of this type our arguments fail, and further investigation
is necessary.

It should be noted that a vanishing elliptic genus does indeed occur
in some important examples. One example arises in $AdS_3 \times S^3
\times T^4$ compactifications \cite{Maldacena:1999bp}. A second
example is  in the MSW conformal field theory with $(0,4)$
supersymmetry, which is dual to an $AdS_3 \times S^2 \times X$
compactification, where $X$ is Calabi-Yau
\cite{Maldacena:1997de,Minasian:1999qn}. In all these cases there is
an extended chiral algebra due to singleton modes.   In such a case
one must take derivatives with respect to $\bar z$ and set $\bar
z=0$ \cite{Cecotti:1992qh,Maldacena:1999bp}. The resulting modular
object is a non-holomorphic generalization of a Jacobi theta
function \cite{deBoer:2006vg,Denef:2007vg}. A similar phenomenon
happens in the analog of the elliptic genus for the \emph{large}
$\CN=4$ superconformal algebra \cite{Gukov:2004fh}.  Of course, the
examples we have just cited are not extremal theories. However,
these examples do suggest that it would be useful to extend the
investigation of extremal theories to the cases of   vanishing
elliptic genera, or $(0,4)$ supersymmetry, or  large $\CN=4$
supersymmetry.


\section{Applications to flux compactifications}
\label{sec:AppFluxCompt}

\newcommand\mpt{M_{pl}^{(3)}}
\newcommand\mpe{M_{pl}^{(11)}}

Flux compactifications of M-theory and string theory have been a
very popular subject of investigation in recent years
\cite{Douglas:2006es,Denef:2007pq}. Unfortunately, these
compactifications are in general very complicated and it is
difficult to be sure that they are valid solutions of string theory
within a controlled approximation scheme. The demonstration of
holographically dual conformal field theories would definitively
settle such difficulties, at least for flux compactifications to
anti-de Sitter spacetimes.  The considerations and techniques of
this paper might put interesting constraints on the allowed spectra
of some classes of flux compactifications, namely compactifications
to $AdS_3$ with a holographically dual $(2,2)$ conformal field
theory. One could imagine, for example, flux compactifications of
M-theory on a suitable Calabi-Yau 4-fold, where one includes $M5$
instanton effects, in order to exclude no-scale compactifications.

The compactifications of greatest interest are those with a small
cosmological constant and a large gap from the ground state to the
Kaluza-Klein scale. These simple aspects of the spectrum already
have implications for the conformal field theory.  If the
cosmological constant is small then the  Brown-Henneaux central
charge $ c = \frac{3}{2 }R\mpt $ is large. This implies
that the level
\be m = \frac{R \mpt}{4} \ee
is large.

Now let us consider the spectrum of the theory. The supergravity
multiplet corresponds to the super-Virasoro descendants. Next, if
$V_8$ is the volume of the Calabi-Yau 4-fold in 11-dimensional
Planck units then
\begin{equation}
[V_8 (\mpe)^8] \mpe = \mpt
\end{equation}
and therefore, $\mpe \sim \mpt$ unless $V_8$ is unnaturally large,
and hence in AdS units, the KK scale is of order $m$. Thus, we
naturally expect a large gap to the primary fields corresponding to
the KK modes.

In addition to the supergravity multiplet and the KK modes there
will typically be other primary fields, for example the moduli
fields, many of which might have acquired masses in the
compactification scheme. Our conjectured bound  (\ref{bound}) might
possibly put  constraints on the masses which the moduli acquire.

%
%
%
%
%
%

It would clearly be of interest to make these considerations more
precise, and moreover to extend them to theories with holographic
duals with only $(1,1)$ supersymmetry. Indeed, one does not expect
generic flux compactifications to lead to $\CN=2$ supersymmetry
since there is no candidate isometry for the $U(1)$ current algebra.
Given the $\CN=1$ supersymmetry one can still form a holomorphic
elliptic genus, but the existence of the Hauptmodul $\hat K$ for $
\Gamma_\theta$ (see eq. (\ref{eq:KhatHaupt}) above) shows  that the
techniques of this paper cannot be used to exclude
compactifications just based on the polar polynomial of the elliptic
genus. Further work is needed to see whether modularity, combined
with other ideas, puts any interesting constraints on the landscape
of three-dimensional AdS compactifications.\footnote{We thank F.
Denef for important discussions that have clarified our
understanding of the prospects for these applications.}

\acknowledgments

We would like to thank C. Vafa for collaboration at an earlier stage
of this project. GM would also like to thank M. Douglas for a past
collaboration on closely related issues. We would like to thank
F.Denef, T. Gannon, S. Kachru, J. Maldacena, J. Manschot, P. Sarnak,
and D. Zagier for useful discussions.

GM and HO thank the organizers of the 37th Paris
Summer Institute on Black Holes, Black Rings and Modular Forms,
which stimulated progress in this work.
SG acknowledges the hospitality of
Institut f\"ur Theoretische Physik, ETH Zurich,
Institute for Advanced Study, Harvard University, Banff Center,
the Aspen Center for Physics, the Simons Workshops in 2005, 2006, and 2007,
where part of this work was carried out.
HO also thanks the Aspen Center for Physics,
the Kavli Institute for Theoretical Physics in Santa Barbara,
Harvard University, the Simons Workshops in 2005 and 2006
in Stony Brook, the Banff International Research Station, the University of Tokyo,
the Galileo Galilei Institute in Florence,
the CERN theory institute, and the Ettore Majorana Centre
for Scientific Culture in Erice, where part of this work was
carried out.

The work of MRG and CAK is supported by the
Swiss National Science Foundation.
GM is supported by DOE grant DE-FG02-96ER40949.
The work of SG and HO is supported in part by
DOE grant DE-FG03-92-ER40701. The work of SG is
also supported in part
by NSF Grant DMS-0635607 and by the Alfred P.
Sloan Foundation. The work of HO is also
supported in part by
NSF grant OISE-0403366, by a Grant-in-Aid for Scientific
Research (C) 20540256 from the Japan Society
for the Promotion of Science, by the 21st
Century COE Visiting Professorship at the University of Tokyo,
by the World Premier International Research Center
Initiative of MEXT of Japan, and by the Kavli Foundation.
Opinions and conclusions expressed here are those of the authors
and do not necessarily reflect the views of funding agencies.

\appendix

\section{Growth properties}\label{s:estimates}

\subsection{Analysis of the constraint for $m$ odd}

 For $m$ odd we have
\be \label{eq:CoeffToEstimateOdd} L=\biggl[ -2i q^{-\frac{m}{4}+
\frac{1}{8}}  \frac{1-q^{1/2}}{1+q^{1/2}}
\frac{\vartheta_3(\tau)}{\eta^3} \vartheta_2^2 \vartheta_4^2
\biggr]_{q^0}   \ . \ee
where $L$ was defined in (\ref{eq:CoeffToEstimate}).  We can
simplify this significantly using the triple product identity
$\vartheta_2 \vartheta_3 \vartheta_4 = 2 \eta^3$. Next, shifting
$\tau\to \tau+1$ (which cannot change the $q^0$ coefficient) we
obtain:
\be  L=4 e^{-i \pi m/2} \left[ q^{-\frac{m}{4}+
\frac{1}{8}}\frac{1+q^{1/2}}{1-q^{1/2}}\vartheta_2 \vartheta_3
\right]_{q^0} \ .\ee
Now use the usual sum formula for $\vartheta_2$
and $\vartheta_3$ to obtain
\be
\vartheta_2 \vartheta_3 =
\sum_{r,s\in \mathbb{Z}} q^{(r-1/2)^2/2 + s^2/2}
= \sum_{r,s\in \mathbb{Z}} q^{(2r-1)^2/8 + (2s)^2/8}
= \sum_{n\in\mathbb{N}_0} B(n) q^{n/8} \ ,
\ee
where $B(n)$ is the number of ways of writing $n$
as a sum of an even and an odd integer squared,
\ie
$n=(2r-1)^2+(2s)^2$ with both $r$ and $s$ integer.
We also observe that the series expansion of the other factor is
\be \frac{1+q^{1/2}}{1-q^{1/2}}= 1 + 2 \sum_{\ell=1}^\infty q^{\ell/2} \ . \ee
Thus the exact result for (\ref{eq:CoeffToEstimate}) is
\be L=  4 e^{-i \pi m/2} \left[ B\left(2m-1\right) + 2
\sum_{\ell=1}^{\frac{2m-1}{4}} B\left(2m-1-4\ell\right) \right]\
.\ee
The dominant contribution comes from the second term. This sum is
precisely equal to all combinations of an odd and an even integer
whose square sum up to a number less or equal to $2m-5$. Now draw a
rectangular lattice whose unit cell is a square with length $2$,
where we shift the lattice by one unit in the $x^1$-direction say,
so that the centers of the cells are at $(x^1,x^2)=(2r-1,2s)$.
Consider the area of all those unit cells for which the
corresponding center point $(2r-1,2s)$ has the property that
$(2r-1)^2+(2s)^2\leq 2m-5$. It follows from elementary geometry that
this area is bigger than the area of the disk with radius
$\sqrt{2m-5}-\sqrt{2}$ (see figure~\ref{figure_modd}). Since
each unit cell has area $4$, it follows that
\be \sum_{\ell=1}^{\frac{2m-1}{4}}
B\left(2m-1-4\ell\right) \geq \frac{1}{4} \, \pi \left( \sqrt{2m-5}
- \sqrt{2} \right)^2 = \frac{\pi}{2} \, m - \pi \sqrt{m-\frac{5}{2}}
- \frac{3}{4} \pi \ . \ee
Thus it follows that $e^{i \pi m/2} L$, which is positive, is
bounded below by
\be e^{i \pi m/2} L \geq 4 \pi m - 8\pi \sqrt{m-\frac{5}{2}} - 6\pi
\ .
\ee

\EPSFIGURE{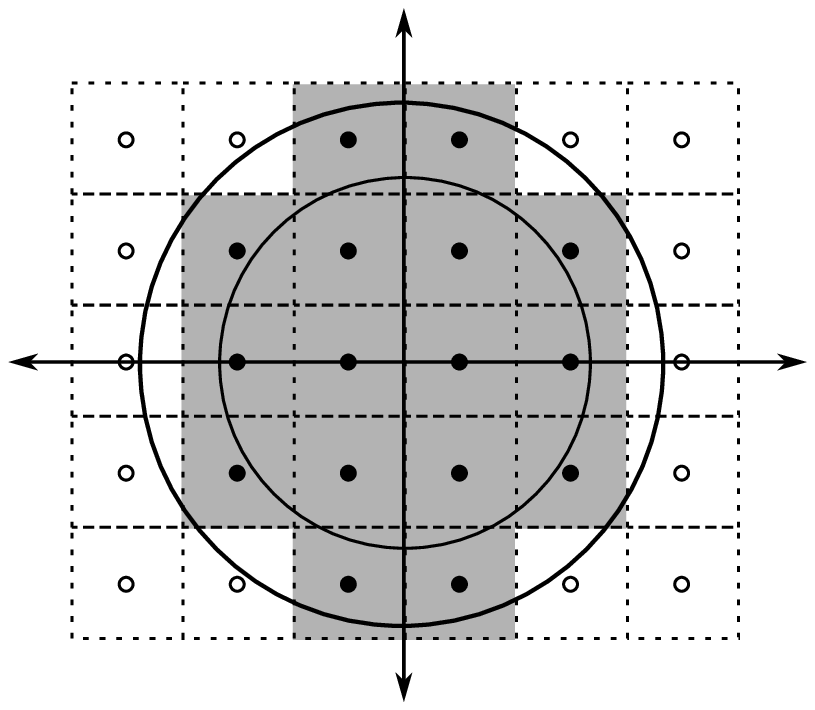,height=6cm,angle=0,trim=0 0 0 0}%
{The grey area is given by those boxes whose centers lie within
the outer circle of radius $\sqrt{2m-5}$. The inner circle has radius
$\sqrt{2m-5}-\sqrt{2}$ and is completely contained in the grey area.
 \label{figure_modd} }


\subsection{Analysis of the constraint for $m= 2\ \mod\ 4\ $}
\label{subsec:m2mod4}
In the case of $m$ odd we saw that $L$ only grew linearly. Since the
original expression contained exponentially growing function such as
$\eta^{-3}$, this means that there had to occur cancellations.
We will now show that for $m = 2\, \mod\, 4$ such cancellation do not
occur, \ie that
\be \label{eq:CoeffToEstimateSimp} L= \biggl[   q^{-\frac{m}{4}+
\frac{1}{8}}\frac{1-q^{1/2}}{1+q^{1/2}}
\frac{\vartheta_3}{\eta^3}(\vartheta_4^4- \vartheta_2^4)
\biggr]_{q^0}   \ee
grows exponentially with $m$.
To this end, use (\ref{eq:logderivtheta})
to write
\be  \left[ q^{-\frac{1}{2}+\frac{1}{8}}\frac{(1-q^{1/2})^2}{1-q}
\left(
    -24\frac{q\partial_q \vartheta_3}{\eta^3}
+ \frac{E_2\vartheta_3}{\eta^3}\right) \right]_{q^N}\ ,
\label{2mod4full} \ee
where $N= m/4-1/2$. The following form of $E_2$ will be useful: \be
E_2(\tau)= 1 - 24 \sum_{k=1}^\infty \sigma_1(k) q^{k} \ , \ee where
$\sigma_1(k)$ is the divisor function.

Let us first consider the second term of (\ref{2mod4full}). We will
show that this is negative and grows exponentially fast with $N$. We
introduce the expansion coefficients of   $\vartheta_3/\eta^3$,
\be \frac{\vartheta_3}{\eta^3} = q^{-1/8}\sum_{n\geq 0} (F_1(n)q^n+
F_2(n)q^{n+1/2})\ . \ee
 From these we obtain the
discrete derivative $(1-q^{1/2})^2\vartheta_3/\eta^3$,
\begin{equation}
q^{-\half + \frac{1}{8}} (1-q^{1/2})^2 \frac{\vartheta_3}{\eta^3} =
\sum_{n\geq 0} \left( K(n) q^n + K'(n) q^{n-1/2}\right)
\end{equation}
with $K(n) = F_2(n) - 2 F_1(n) + F_2(n-1)$, and, including $E_2$,
\begin{equation}
q^{-\half + \frac{1}{8}}E_2 (1-q^{1/2})^2 \frac{\vartheta_3}{\eta^3}
= \sum_{n\geq 0} \left( \hat K(n) q^n +\hat K'(n) q^{n-1/2}\right)
\end{equation}
with
\be \hat K(n) = K(n) - 24 \sum_{s=1}^n \sigma_1(s) K(n-s)\ . \ee
Finally, the desired second term of (\ref{2mod4full}) is $\sum^N
\hat K(n) $. It will therefore suffice to show that $\hat K(n)$
grows exponentially and is negative for large $n$.

To examine the large $n$ behavior we begin with the Rademacher
expansions for $F_1(n)$ and $F_2(n)$. These are summarized in
appendix~\ref{s:Rademacher} with the result that
\begin{eqnarray*}
F_1(n) &=& (8n)^{-5/4}e^{\pi\sqrt{2n}}\left( 1 - \frac{15+\pi ^2}{8
\sqrt{2} \pi }n^{-1/2} + \frac{105+10 \pi ^2+\pi ^4}{256 \pi ^2
}n^{-1} +
\CO(n^{-3/2})\right)\ , \\
F_2(n) &=& (8n)^{-5/4}e^{\pi\sqrt{2n}}\left( 1+\frac{3 \left( \pi
^2-5\right)}{8 \sqrt{2} \pi} n^{-1/2} +\frac{3 \left(35-10 \pi ^2+3
\pi^4\right)}{256 \pi ^2} n^{-1} + \CO(n^{-3/2})\right)\ .
\end{eqnarray*}
{}From this we compute the discrete derivative:
\be K(n) =  \pi^2 (8n)^{-9/4}e^{\pi\sqrt{2n}}(1 + O(n^{-1/2})) \ .
\label{Kasym}
\ee
Note the exponential growth with $n$.  Now write
\be \label{eq:estkhat}\hat K(n) = K(n) - 24   K(n-1) - 24 S
 \ee
with $S:=\sum_{s=2}^n \sigma(s) K(n-s)$. It is straightforward to
see that the sum   $S$ is positive definite for large $n$: first
note that because of (\ref{Kasym}) $K(n)$ is negative for at most
finitely many $n$. Since $K(n)$ grows exponentially and $\sigma(s)$
only grows like   $\sigma(s) \sim e^\gamma s \ln \ln s$,  where
$\gamma$ is the Euler-Mascheroni constant \cite{HardyWright}, it
follows that the first terms of the sum dominate the (potentially
negative) terms at its tail. The first two terms on the RHS of
(\ref{eq:estkhat}) clearly grow like $-23 \pi^2 (8n)^{-9/4} e^{\pi
\sqrt{2n}}$, hence $\hat K(n)$ is negative and exponentially growing
for large $n$. Therefore the same is true for $\sum^N \hat K(n)$.

In the analysis of the case $m = 0\, \mod\, 4$ below we will show
that the first term of (\ref{2mod4full}) is negative, so that
there can be no cancellations between the two. We thus conclude that
(\ref{2mod4full}) grows exponentially.

\subsection{Analysis of the constraint for $m= 0\ \mod\ 4\ $}
Define
\begin{eqnarray}
R_1 &=& \left[ 2\, q^{1/2} \, \frac{(1-q^{1/2})^4}{(1-q)^3}
\frac{
  \vartheta_3}{\eta^3}\right]_{q^{\frac{m}{4}-\frac{1}{8} }}
\label{R1full}  \\
R_2 &=&  \left[ \frac{(1-q^{1/2})^2}{1-q} \frac{q \partial_q
\vartheta_3}{\eta^3}\right]_{ q^{\frac{m}{4}-\frac{1}{8} }} \ .
 \label{R2full}
\end{eqnarray}
We shall show that for large enough $m$ both $R_1$ and $R_2$ are
positive. Consider first $R_2$. Note that the only negative
coefficients that can appear are due to the factor $(1-q^{1/2})^2$.
It will suffice to show that the coefficients
\be
\left[ \frac{(1-q^{1/2})^2}{(1-q)^3(1-q^2)^3}\, q \partial_q
  \vartheta_3 \right]_{q^N}
\label{exprC2}
\ee
are positive for $N$ large enough. We have dropped the factor of
$(1-q)^{-1}$ and included only the first two factors of $\eta^3$,
which will turn out to be sufficient to ensure positivity. Defining
\be
\frac{1}{(1-q)^3(1-q^2)^3}=\sum_{n=0}^\infty b(n)q^n \ ,
\ee
it is straightforward to calculate
\be
b(n) = \left\{\begin{array}{cc}
  \frac{1}{1920}(2+n)(4+n)(6+n)(8+n)(5+2n) \qquad  & n\ \textrm{even} \\
\frac{1}{1920}(1+n)(3+n)(5+n)(7+n)(13+2n) \qquad  & n\ \textrm{odd.}
\end{array}
\right.  \label{bn}
\ee
Note in particular that
\be
b(n) = \frac{n^5}{960}+\frac{3 n^4}{128}+\frac{19 n^3}{96} + \CO(n^2)\ .
\ee
We now want to calculate the coefficients $p(N)$ of
\be \frac{1}{(1-q)^3(1-q^2)^3}\, q \partial_q \vartheta_3 = \sum_{N
\in
  \frac{1}{2}\mathbb{N}} p(N)q^N\ .
\ee
We need to distinguish the cases $N \in \mathbb{N}$ and $N \in
\mathbb{N} + \frac{1}{2}$:
\begin{eqnarray}
N \in \mathbb{N} &:& p(N,K)= \sum_{s=0}^K b(N-2s^2)\, 4s^2
\label{exprpN2} \\
N \in \mathbb{N} + \frac{1}{2} &:& p(N,K) = \sum_{s=0}^K b(N-
(2s+1)^2/2) \, (2s+1)^2 \label{exprpN}
\end{eqnarray}
In principle, the upper bound $K$ is given by the requirement that the
argument of $b$ be non-negative, and its explicit expression will
involve some floor function
of a square root of $N$.
For the moment, we will leave $K$ as an auxiliary integer
parameter. One can then evaluate the sums explicitly to obtain
polynomials in
both $N$ and $K$, again distinguishing
the cases $N$ odd and $N$ even. As the resulting expressions are
rather lengthy,
we refrain from writing them down explicitly.
To determine the $N$th coefficient of (\ref{exprC2}), we then need to
evaluate
\be
p(N,K_1)-2p(N-1/2,K_2) + p(N-1,K_3)\ . \label{disderC2}
\ee
In principle, we would now have to determine the exact values of
$K_i$, which are complicated step functions of $N^{1/2}$. For our
purposes however it is enough to know their leading behavior. In
particular,
we know that $K_i = \sqrt{\frac{N}{2}}-\epsilon_i$, where $0\leq
\epsilon_i < 2$, so that $\epsilon_i$ is of order one.
We then obtain for (\ref{disderC2}) the expression
\be
\frac{N^{9/2}}{1890 \sqrt{2}} + \CO(N^{7/2}) \ .
\ee
Note that this holds for all $N \in \frac{1}{2} \mathbb{N}$. (Hence,
our estimates can also be applied to the analysis of section
\ref{subsec:m2mod4}.) This shows that the leading term has a
positive coefficient and that it is independent of the $\epsilon_i$,
which only appear in the subleading terms. This then shows that
(\ref{exprC2}) has positive coefficients for $N$ large enough.

Note that for low values of $N$ the coefficients of (\ref{exprC2})
can still be negative. To complete the argument, we thus have to
show that after convolution with the remaining factors in
(\ref{R2full}) the potentially negative coefficients for $N<N_0$
cannot render negative the coefficients at arbitrarily large $N$. To
see this, note that it follows from the Rademacher expansion that,
for any set of positive integers $ a_1, \dots, a_k$, the Fourier
coefficients of
\be (1-q)^{a_1} (1-q^2)^{a_2}\cdots (1-q^k)^{a_k} \eta^{-3} \ee
will have the asymptotic behavior $\sim n^p e^{\pi \sqrt{2n }}$. For
example in   Appendix B we show that for the case of interest,
$(1-q)^3 (1-q^2)^3 \eta^{-3}$ the leading asymptotics is given by
\be \frac{\pi ^6}{8 \sqrt{2}} n^{-9/2}e^{\pi\sqrt{2n}} \ . \ee
We approximate the convolution sum as the integral
\be \int^N\!\!\! ds\, s^{9/2} (N-s)^{-9/2}\, e^{\pi \sqrt{2 (N-s)}}\
. \ee The position of the saddle point of this integral  grows as
\be
s_0 \sim N^{1/2}\ .
\ee
This means that for $N$ large enough the
contribution of the negative coefficients around $s \sim 1$ will be
negligible, so that the total coefficient is positive.

Turning to $R_1$, we need to consider
\be
(1-q)^{-3} (1-q^2)^{-3}(1-q^3)^{-3}(1-q^4)^{-3} =
\sum_{n=0}^\infty \tilde b(n)q^n \ .
\ee
A straightforward, but somewhat tedious calculation then gives
expressions similar to (\ref{bn}) whose explicit forms depend on $n\
\mod\ 12$. Again, the leading terms are independent of this, so
that we can write
\be
\tilde b(n) = \frac{n^{11}}{551809843200} + \frac{n^{10}}{3344302080}
+ \frac{29\ n^9}{1337720832} + \frac{5\ n^8}{5505024} + \frac{16949\
  n^7}{696729600} + \CO(n^6)\ . \label{tildeb}
\ee
We can now
define $\tilde p(N,K)$ analogously to (\ref{exprpN2}), (\ref{exprpN})
and evaluate
\be
\tilde p(N,K_1)-4 \tilde p(N-1/2,K_2) + 6 \tilde p(N-1,K_3) - 4\tilde
p(N-3/2,K_4) + \tilde p(N-2,K_5)\ ,
\ee
which leads to \be \frac{N^{15/2}}{1751349600 \sqrt{2}} +
\CO(N^{13/2})\ . \ee Since sums of terms of order $n^6$  give
contributions of at most $N^7$, this also shows that it was
sufficient to consider (\ref{tildeb}) only up to $n^6$. The
coefficients of the truncated $\eta^{-3}$ expansion grow as in
(\ref{eq:TruncFour}), and the   rest of the argument is then
completely analogous to the case of $R_2$.

\section{Rademacher expansions}\label{s:Rademacher}

The proofs in appendix~\ref{s:estimates} require some
asymptotic expansions for coefficients of some modular forms. We
collect these here.

First, we apply the expansion to the modular vector

\begin{eqnarray}
f_1 & =& \half \frac{\vartheta_3+ \vartheta_4}{\eta^3} =
q^{-1/8}\sum_{n=0}^\infty F_1(n) q^{n}\\
f_2 & =& \half \frac{\vartheta_3 - \vartheta_4}{\eta^3} =
q^{3/8}\sum_{n=0}^\infty F_2(n) q^{n}\\
f_3 & =&  \frac{\vartheta_2}{\eta^3}=  \sum_{j=0}^\infty F_3(n)
q^{n}\ .
\end{eqnarray}
We have weight $w=-1$, the representation is manifest for $T$, and
for $S$ it is computed from
\begin{eqnarray}
f_1(-1/\tau) &
= & (-i \tau)^{-1} \half (f_1 + f_2 + f_3)(\tau) \\
f_2(-1/\tau) &
=& (-i \tau)^{-1} \half (f_1 + f_2 - f_3)(\tau) \\
f_3(-1/\tau) &
=& (-i \tau)^{-1}  (f_1 - f_2  )(\tau)\ . \\
\end{eqnarray}
We now have convergent expansions
\begin{eqnarray}
F_1(n) &
= & \frac{\pi}{8} (n-1/8)^{-1} I_2( 4 \pi
\sqrt{\frac{1}{8}(n-\frac{1}{8})})+\CO(e^{2\pi\sqrt{n/8}})  \\
F_2(n) &
= & \frac{\pi}{8} (n+3/8)^{-1} I_2( 4 \pi
\sqrt{\frac{1}{8}(n+\frac{3}{8})})+\CO(e^{2\pi\sqrt{n/8}}) \\
F_3(n) &
= & \frac{\pi}{8} (n )^{-1} I_2( 4 \pi \sqrt{\frac{1}{8} n })
+\CO(e^{2\pi\sqrt{n/8}})\ .
\end{eqnarray}
Now use
\begin{equation}
I_\nu(x) \sim \frac{1}{\sqrt{2\pi x}} e^x \left(1 -
\frac{4\nu^2-1}{8x} + \frac{(4\nu^2-1)(4\nu^2-9)}{128 x^2} + \cdots
\right)
\end{equation}
for $x\to +\infty$ to get
\begin{equation}\label{eq:RadAns1}
F_1(n) = (8n)^{-5/4} e^{ 4\pi \sqrt{\frac{n}{8}} } \Biggl(1-
\frac{\pi^2 + 15}{8 \sqrt{2}\pi} \frac{1}{n^{1/2}} + \frac{\pi^4 +
70 \pi^2 + 105}{256\pi^2}\frac{1}{n} + \cdots \Biggr)
\end{equation}
\begin{equation}\label{eq:RadAns2}
F_2(n) = (8n)^{-5/4} e^{ 4\pi \sqrt{\frac{n}{8}} } \Biggl(1+
\frac{3(\pi^2 - 5)}{8 \sqrt{2}\pi} \frac{1}{n^{1/2}} +
\frac{3(3\pi^4 - 70 \pi^2 + 35)}{256\pi^2}\frac{1}{n} + \cdots
\Biggr)\ .
\end{equation}

We also need the asymptotic expansion of functions that are obtained
from $\eta^{-3}$ by dropping the first few factors in the product
formula. Defining
\begin{equation}
\eta^{-3} = q^{-1/8} \sum_n p_3(n)q^n
\end{equation}
(with $p_3(n)=0$ for $n<0$ ), we have the Rademacher formula
\be \label{p3nRademacher}
p_3(n) = 2\pi (8n-1)^{-5/4} I_{3/2}(\pi \sqrt{2(n-1/8)}) +
\CO(e^{\pi\sqrt{n/2}})\ .
 \ee
Note that the Bessel function is elementary
\be I_{3/2}(x) = \frac{2}{\sqrt{2\pi x}}(\cosh x - \frac{\sinh
x}{x}) \ .
\ee
Define
\begin{equation}
(1-q)^3 (1-q^2)^3 \eta^{-3} = q^{-1/8} \sum_n \hat p_3(n)q^n \ ,
\end{equation}
which is a kind of sixth-order discrete derivative:
\begin{multline}
 \hat p_3(n) = p_3(n) -3 p_3(n-1) + 8 p_3(n-3) \\
- 6 p_3(n-4) - 6
p_3(n-5) + 8 p_3(n-6) - 3 p_3(n-8) + p_3(n-9)  \ .
\end{multline}
Substituting the asymptotic expansion (\ref{p3nRademacher}) one finds
after some algebraic manipulations
\be \hat p_3(n) = \left(\frac{\pi ^6}{8 \sqrt{2}}n^{-9/2} +
\CO(n^{-5})\right) e^{\pi \sqrt{2n}} \ .
 \ee
Similarly, the coefficients
\begin{equation}
(1-q)^3 (1-q^2)^3 (1-q^3)^3 (1-q^4)^3\eta^{-3} = q^{-1/8} \sum_n
\tilde  p_3(n)q^n
\end{equation}
have leading asymptotics
\be\label{eq:TruncFour} \tilde p_3(n) \sim \left(\frac{27
\pi^{12}}{\sqrt{2}} n^{-15/2 }+\CO(n^{-8})\right) e^{\pi \sqrt{2n}} \ .
\ee
%


\end{document}